\documentclass{emulateapj}
\usepackage{ulem}
\usepackage{xcolor}

\usepackage{graphicx}

\shorttitle{X-ray and Optical Emission in Cas A}

\shortauthors{Patnaude \& Fesen}

\begin{document}

\title{A Comparison of X-ray and Optical Emission in Cassiopeia A}

\author{Daniel J.~Patnaude\altaffilmark{1} and
Robert A.~Fesen\altaffilmark{2}}
\altaffiltext{1}{Smithsonian Astrophysical Observatory, 
Cambridge, MA 02138, USA}

\altaffiltext{2}{6127 Wilder Lab, Department of Physics \& 
Astronomy, 
Dartmouth College, Hanover, NH 03755, USA}

\begin{abstract}

Broadband optical and narrowband \ion{Si}{13} X-ray images of the young
Galactic supernova remnant Cassiopeia A (Cas A) obtained over several decades
are used to investigate spatial and temporal emission correlations on both
large and small angular scales. The data examined consist of optical and near
infrared ground-based and {\it Hubble Space Telescope} images taken between
1951 and 2011, and X-ray images from {\it Einstein}, {\it ROSAT}, and {\it
Chandra} taken between 1979 and 2013. We find weak spatial correlations between
the remnant's X-ray and optical emission features on large scales, but several
cases of good optical/X-ray correlations on small scales for features which
have brightened due to recent interaction with the reverse shock. We also find
instances where: (i) a time delay is observed between the appearance of a
feature's optical and X-ray emissions, (ii) displacements of several arcseconds
between a feature's X-ray and optical emission peaks and, (iii) regions showing
no corresponding X-ray or optical emissions.  To explain this behavior, we
propose a highly inhomogeneous density model for Cas A's ejecta consisting of
small, dense optically emitting knots (n $\sim 10^{2-3}$ cm$^{-3}$) and a much
lower density (n $\sim 0.1 - 1$ cm$^{-3}$) diffuse X-ray emitting component
often spatially associated with optical emission knots.  The X-ray emitting
component is sometimes linked to optical clumps through shock induced mass
ablation generating trailing material leading to spatially offset X-ray/optical
emissions.  A range of ejecta densities can also explain the observed
X-ray/optical time delays since the remnant's $\approx 5000$ km s$^{-1}$
reverse shock heats dense ejecta clumps to temperatures around $3 \times
10^{4}$ K relatively quickly which then become optically bright while more
diffuse ejecta become X-ray bright on longer timescales.  Highly inhomogeneous
ejecta as proposed here for Cas A may help explain some of the X-ray/optical
emission features seen in other young core collapse SN remnants. 

\end{abstract}


\section{Introduction}

With an estimated undecelerated explosion date around 1680
\citep{thor01,fesen06}, Cassiopeia A (Cas A) is one of the youngest known
Galactic supernova remnants (SNRs).  It is also one of the few historic
remnants with a secure supernova (SN) subtype through the detection 
of optical and infrared light echoes of its initial supernova outburst
\citep{krause08,rest08,rest11,Besel12}.  

Optical spectra of its light echo show Cas A to be the remnant of a
core-collapse Type IIb supernova event with an optical spectrum at maximum
light similar to SN 1993J in M81. As indicated by the slow dense wind into
which Cas A's forward shock is expanding \citep{chevalier03,hwang09}, the Cas~A
progenitor was probably a red supergiant with a mass of 15--25M$_{\sun}$ that
may have lost much of its hydrogen envelope to a binary interaction
\citep{young06,hwang12}.

Viewed in X-rays, the remnant consists of a bright, line-emitting shell arising
from reverse shocked ejecta rich in O, Si, S, Ar, Ca, and Fe \citep{fabian80,
markert83,vink96,hughes00,willingale02,willingale03,hwang03,laming03}. Small
knots and filamentary regions of X-ray emitting ejecta have been
observed to change in intensity and structure over time, indicating the
location of recently shocked, ionizing ejecta \citep{patnaude07}. Exterior to
this shell are faint X-ray filaments which mark the current location of the
SNR's $\simeq$ 5000 km s$^{-1}$ expanding forward shock
\citep{delaney04,patnaude09a}. This emission is largely nonthermal but can
include faint line emission from shocked circumstellar material
\citep[CSM;][]{araya10}. This outer nonthermal X-ray emission
is fading with time \citep{patnaude11} while the bulk of the remnant's
bright thermal emission arising from shocked ejecta has remained relatively
steady over the last few decades.

The remnant's optical and infrared emissions trace the location of
its denser debris ($\geqq 10^{3}$ cm$^{-3}$; \citealt{Hurford96,Fesen2001})
which in some places is co-spatial with lower density and more
diffuse X-ray emitting material. The bulk of Cas~A's optical and near-infrared
emission consists of a V$_{\rm r}$ =  $-4000$ to $+6000$ km s$^{-1}$
expanding shell of knots, condensations, and filaments which lack any H$\alpha$
emission \citep{ck78,ck79,Reed1995,delaney10,MF2013}. A few dozen
semi-stationary condensations known as QSFs which do exhibit strong H$\alpha$
and [\ion{N}{2}] $\lambda\lambda$6548, 6583 emissions with prevalently negative
radial velocities of 0 to $-250$ km s$^{-1}$ appear to be pre-SN, circumstellar
mass-loss material \citep{vdbD70,vdb71b,vandenbergh85,Reed1995}.

In the picture of an inhomogeneous SN debris field with a range of ejecta
densities and component dimensions, strong and widespread correlations between
low density X-ray emitting material and small dense optically emitting knots
are not expected; optical emission arises from gas with electron temperatures of
around $\sim$ 30,000 K while X-rays arise from material shock heated to several
million degrees K.  Indeed, one generally sees only a weak spatial correlation
of bright fine scale X-ray and optical features in images of Cas~A taken at
similar epochs \citep{laming03}.  But there are exceptions.  For example,
while noting only broad optical/X-ray emission coincidences in the remnant's
northern regions in the 1979 {\sl Einstein} image,
\cite{fabian80} cited an especially good correspondence between optical, X-ray, and radio
emission in the bright optical Filament 1 of \citet{BM54}. 

However, in places where dense optical emitting knots are embedded within or
associated with a much lower density component there might be a close optical/X-ray
emission correlation. The resemblance of the remnant's morphology in
the optical and Si and S X-ray emission lines is suggestive of at least 
some spatial
coincidence \citep{Hwang00}. Furthermore, indications of ejecta knot mass
stripping in Cas A have been reported \citep{Fesenetal2001,Fesen11} and such
trailing material, if of the right density and mass, could lead to detectable
associated X-ray emission downstream.  Lastly, different excitation
timescales between optical and X-ray emissions might lead to discernible time
lags between the onset of strong optical and strong X-ray emissions. 

Here we present the results of an investigation into X-ray and optical
correlations that may be related to such factors through a comparative survey
of Cas A's optical and X-ray emission evolution over the last 30 years.  The
observations and results are described in $\S$2 and $\S$3, with a modeling
analysis described in $\S$4.  A summary of our findings and conclusions is
given in $\S$5.

\section{Observations and Data Reduction}

\subsection{X-Ray Imaging Data}

The earliest X-ray images of Cas~A with angular resolution capable of resolving
features at scales $\sim$ 5$\arcsec$ were obtained in 1979 with the {\sl
Einstein} satellite and its High Resolution Imager (HRI).  Details of the {\sl
Einstein} telescope and HRI detector can be found in \citet{giacconi79}.  The
{\sl Einstein} HRI has little spectral resolution over the 0.1--4.5 keV
bandwidth of the telescope, and hence we used the broadband image without
filtering any pulse height channels. 

Cas~A was not imaged again at high angular resolution until 1995--1996, 
with the {\sl
ROSAT} HRI, in the 0.2--2.4 keV band \citep{koralesky98}. Like the {\sl
Einstein} observations, we do not attempt any energy filtering of
the {\sl ROSAT} HRI data. Due to the high column density towards
Cas A, the X-ray emission in this bandpass is dominated by thermal emission
from \ion{Si}{13} at $\simeq$ 1.85 keV. 

More recently, Cas~A has been extensively observed with the {\it Chandra} X-ray
Observatory using the ACIS-S CCD detector. We summarize those observations (as
well as the prior X-ray observations) used in this paper in
Table~\ref{tab:obs}. Since we are looking for correlations between Cas A's
X-ray and optical emissions and because a majority of archival optical images
were particularly sensitive to 
[\ion{S}{2}] $\lambda\lambda$6716,6731 emissions we
focused our attention on the He-like ion emission of silicon in the {\it
Chandra} X-ray data which should emphasize S + Si-rich ejecta.

We reprocessed all of the {\it Chandra} ACIS observations of Cas A using 
{\it ciao}
4.4.1 and CALDB 4.5.1 and then created exposure corrected images using the 
{\it ciao}
script {\tt fluximage}.  Continuum image maps for the energy ranges 
1.6--1.7 keV
and 1.95--2.05 keV were generated, and we used them to estimate 
the continuum emission under
the He--like silicon triplet. We then computed an exposure corrected image for
\ion{Si}{13} from 1.75--1.94 keV (centered on 1.85 keV) and subtracted the
continuum emission image from that. 
This produces a map of emission that is mostly from
\ion{Si}{13}. We did this for each {\it Chandra} observation listed in
Table~\ref{tab:obs}.

The resulting \ion{Si}{13} X-ray images were examined for time variable
features in the form of the emergence or disappearance of individual knots and
filaments. While there is evidence for variability in X-ray  emission over
short time scales \citep{patnaude07}, here we investigated the time when
certain X-ray knots first became visible in order to compare with the formation
of coincident optical emission, where present. 

For comparisons between {\it Einstein}, {\it ROSAT}, and {\it Chandra}
observations we required a different strategy. Since we are unable to energy
filter the {\it Einstein} and {\it ROSAT} observations, we chose to
approximately match the {\it Chandra} observations to those HRI observations by
energy filtering the ACIS-S data between 0.1--4.5 keV. However, the {\it
Chandra} response below 0.3 keV is poorly characterized, so we only filter to
0.3 keV, and in any event low energy X-rays and XUV emission from Cas A are
absorbed by a high column density (i.e., low energy X-rays from Cas A do not
contribute much to the overall detected flux).

Several regions or features which showed significant flux variability using the
1979 {\it Einstein} image, the 1995 {\it ROSAT} image, and the calibrated 2000,
2004, 2007, 2010, and 2013 {\it Chandra} images (energy filtered between
0.3--4.5 keV) were identified and investigated. Using 5$\arcsec$ radius
apertures, we extracted total counts from each data set for each region and
compute the detector count rate, using the exposure times listed in
Table~\ref{tab:obs}.  \citet{hwang12} presented maps of the fitted X-ray
temperature and absorbing column towards Cas A. For the purposes of this study,
we assume that on average, the column towards Cas A is N$_{\mathrm{H}}$ $=$
1.5$\times$10$^{22}$ H atoms cm$^{-2}$, and $k_{\mathrm{B}}$T $=$ 1.8 keV.
Using \texttt{WebPIMMS}, we converted the detector count rates into absorbed
source fluxes, assuming a solar abundance APEC model with the above parameters.
The results for five features discussed in \S~\ref{sub:smallscale} are 
listed in
Table~\ref{tab:xrayflux}. We note that while the APEC model does not 
accurately reflect the conditions we are investigating (it being a model
for a plasma in collisional ionization equilibrium), we are using it to
compute flux changes which give rise to the observed changes in count 
rate.

\subsection{Optical Image Data}

For optical observations of Cas~A, we have made use of the extensive archival 
collection of Palomar Observatory 200 inch (5m) photographic plates taken 
of the
remnant between 1951 and 1989 and made available though the assistance
of S. van den Bergh.  Added to this data set, are several more recent 
CCD images of
Cas A taken with the MDM 1.3m and 2.4m telescopes obtained between 1992 and
1999 along with {\sl Hubble Space Telescope} ({\it HST}) images of the remnant
obtained from 2000 to 2011 (see Table~\ref{tab:obs}).

\subsubsection{Palomar Plates: 1951--1989}

The majority of the optical imaging data are the Palomar 5m images, which
comprise the earliest and deepest images taken of Cas A. These form a unique
image set recording the remnant's optical appearance soon after its 1948
discovery via radio observations \citep{Ryle1948} and before the advent of high
efficiency electronic detectors in the 1980s. These plates, $5 \times$ 7
inches in size, were all taken at the Palomar 5m prime focus usually employing the
Ross corrector plate and have a plate scale of $0.11\arcsec$ per mm. Filter and
photographic emulsion details for each plate used or examined for this study
are listed in Table~\ref{tab:obs}.

These photographic plates, after being manually cleaned of surface dust, 
were digitized
using the high-speed digitizing scanning machine DASCH
\citep{Simcoe2006,Laylock2010} at the Harvard--Smithsonian 
Center for Astrophysics.  This
scanner consists of a precision (0.2 $\mu$m) X-Y table, 
a red LED array, and a 4k
$\times$ 4k CCD camera, and is capable of scanning two 8 $\times$ 
10 inch plates in
less than one minute.  It can record the full dynamic range of the photographic
negative (plate), and with its camera pixels just $11\mu$m in size is able to
record plate details finer that the emulsion grains.  Output is a set of
overlapping frames which are 
then mosaiced together to form a single image of
each frame.  Approximate world coordinates (WCS) were determined for each plate
using astrometric solutions done with an automated procedure using software
routines in WCSTools \citep{Laylock2010}. More accurate WCS image coordinates
were subsequently applied using USNO UCAC3 catalog stars \citep{Zach2010} 
with a resulting estimated accuracy of $0\farcs4$ near plate center. 

\subsubsection{Photometric Flux Measurements}

Although the DASCH scanner is capable of generating accurate photometric data
from photographic plates, we have not attempted to flux calibrate these images
to the CCD images in our data set. This was viewed as an 
enormous undertaking (if even possible) and well beyond the 
scope and needs of this study. Aside from the 
lack of calibration plates taken from the same emulsion plate
batches used for each of these images, the heterogeneous variety of
photographic plate emulsions used over several decades (1951--1989), different
chemical developers, development times and procedures means that the
`characteristic curve' of log(exposure) versus log(plate density) is 
likely to be different for every single plate (for further details see
\citealt{Laylock2010}).

Apart from the photographic characteristic curve issue, calibrated optical
magnitudes of the remnant's emission line ejecta knots off the Palomar plates
using a catalog of known magnitudes by assuming a non-linear function like that
done for the similarly scanned Harvard College Observatory plate collection
was also deemed not possible \citep{Laylock2010}. The Palomar plates taken of
Cas~A were obtained through a variety of uncalibrated broadband filters, and
this taken together with the emission line nature of the remnant's optical
flux, i.e., unknown relative and possibly variable [\ion{O}{1}], [\ion{O}{2}],
[\ion{S}{2}], and [\ion{Ar}{3}] emission line strengths within these filter
passbands, greatly limits the meaning of the resulting photographic density to
optical flux conversion.  

The problem of measuring the equivalent of broad passband BVRI or Sloan filter
optical magnitudes of Cas~A's emission knots even applies, although to a lesser
extent, to modern CCD images.  The remnant's optical emission spectrum varies
greatly across filaments and knots \citep{Hurford96} and changes in the
relative line strengths for individual ejecta knots have been
observed, especially for ejecta rapid flux changes such as recently shocked
ejecta knots \citep{Morse2004,Fesen11}.  Consequently, we have limited our
optical/X-ray comparison survey only to those few coincident X-ray and optical
emission knots or regions which exhibited obvious changes in flux.

Given these severe constraints, our estimated optical flux changes using the
Palomar plates are limited only to gross changes in optical brightness for
specific regions or features.  Because magnitude values for late-type stars
m$_{pg}$ = 11.5 to 17.3) in the Cas A field  made by \citet{vdb71a} were
saturated on the Palomar plates we calibrated these images using
USNO A2 red R2 magnitudes for a set of stars around the remnant in the
magnitude range of 18.5 to 19.5.  For red emulsion images using RG2, RG630, and
RG645 filters, the internal consistency was a relatively modest $\pm 0.25 $
mag given the uncalibrated plate material.

However, given the highly inhomogeneous optical image data, namely archival
Palomar plate material, R band CCD images, and {\sl HST} images taken with
three different detectors and filter combinations (WFPC2 W675F, ACS F625W +
F775W, and WFC3 F098M), we were unable to correlate optical fluxes of specific
regions with meaningful accuracy. The basic problem is two-fold: 1) the
different data sets employed different detectors having different wavelength
sensitivities, and 2) a wide range of filters were used which either do not overlap
each other, cover the same set of important emission lines, or have similar
transmission at these line emission wavelengths. This situation made an
accurate assessment of a feature's optical flux changes not possible. Instead
we present optical images of specific regions of interest where a qualitative
assessment of a feature's variability can be made in relation to other nearby
remnant emission regions.

\subsubsection{Blue vs. Red Images}

For our comparison study of Cas A's X-ray and optical emission structures, we
chose to concentrate on red passband optical images.  This decision was driven
in part by the greater quality and quantity of Palomar red plates of Cas~A (see
Table~\ref{tab:obs}) and the increased remnant's 
structure when viewed at longer
wavelengths due to more red line emissions than in the blue. 
Cas A's blue optical line emissions are dominated by [\ion{O}{3}]
$\lambda\lambda$4959,5007 emission, whereas at longer wavelengths the remnant's
emissions arises from a larger number of strong emission lines including
[\ion{O}{1}] $\lambda\lambda$6300,6364, [\ion{S}{2}] $\lambda\lambda$6716,6731,
[\ion{Ar}{3}]$\lambda$7135, and [\ion{O}{2}]$\lambda\lambda$7319,7330.

\begin{figure}
\includegraphics{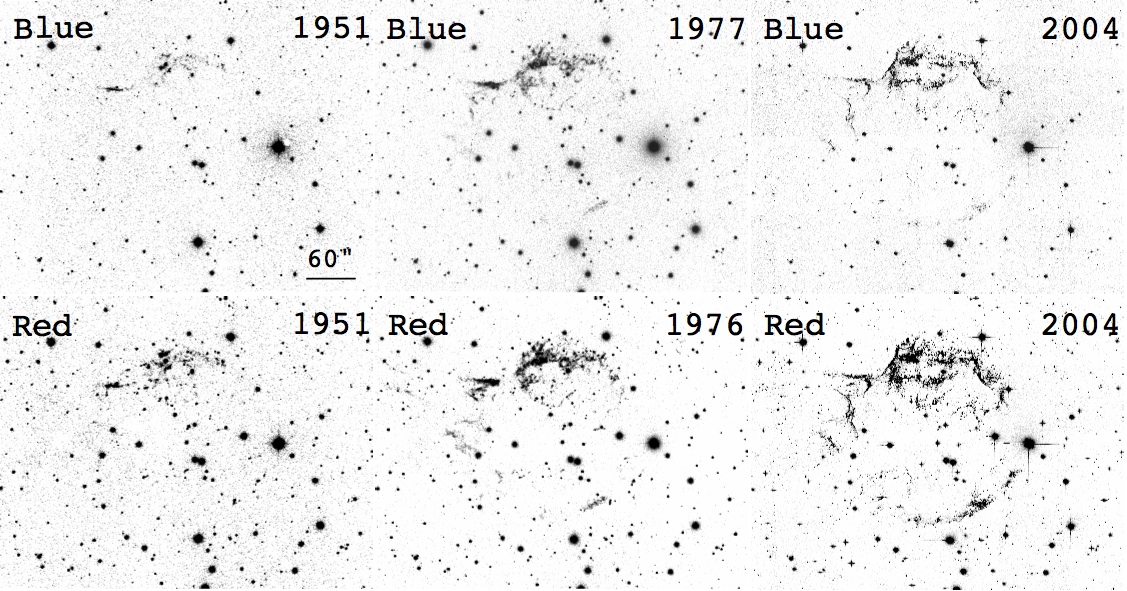}
\caption{
Comparisons of blue and red broadband optical images of Cas A.  Top panels show
Cas~A's optical appearance from 1951 to 2004 in blue light and mainly sensitive  
to [\ion{O}{3}] $\lambda\lambda$4959,5007 emission. Images shown are the 1951 and 1977 Palomar
5m plates PH520B and PH7445, respectively, along with a December 2004 F475W ACS HST image. 
 Lower panels show Cas A over this same period of
time in broadband red images dominated by [\ion{S}{2}]
$\lambda\lambda$6716,6731,
[\ion{O}{1}] $\lambda\lambda$6300,6364 and [\ion{O}{2}]
$\lambda\lambda$7320,7330 line emissions.  The 1951 and 1976 Hale 5m images are plates PH563B
and PH7252vB while the HST image is a co-add of December 2004 ACS F625W and
F775W images. }
\label{fig:blue_vs_red}
\end{figure}

As seen in Figure~\ref{fig:blue_vs_red}, while the remnant's structure changes
substantially between 1951 and 2004 and bright blue and red structures are
consistent with one another, there are significant differences
especially along the southern limb.  Even in the northern regions where they
are similar in gross morphology, remnant features are brighter and better
detected in the red. Such differences are also due in part to the considerable
interstellar reddening toward the remnant 
($A_{V}$ =  4--6 mag; \citealt{Hurford96,hwang12}) 
which greatly weakens the remnant's blue emission
features.  For these reasons, we focused most of our attention on the more
complete data set of red optical images when comparing the remnant's optical
and X-ray emissions.

\section{Results}
\label{sec:results}

Spatial correlations between Cas A's optical and X-ray emission are generally
not strong. Large scale emission correlations that do exist are mainly confined
to the northern and northeastern regions \citep{laming03}. This lack of strong
correlation may be due in part to the relatively high and varied extinction
along the line of sight to Cas A, especially along its western limb
\citep{eriksen09,hwang12}. We have attempted to mitigate some of the effects of
this absorption by using red and near infrared images sensitive to the
remnant's strong [\ion{S}{2}] $\lambda\lambda$6716, 6731 and [\ion{S}{3}]
$\lambda\lambda$9069, 9531 line emissions and by examining the remnant's X-ray
emission from \ion{Si}{13} at $\sim$ 1.8 keV which is not strongly affected by
interstellar absorption. 

In addition, by focusing on emission from silicon (in X-rays) and sulfur (in
optical), one is sampling ejecta with like abundances, as these two elements
are both synthesized during explosive oxygen burning. Using sulfur as a proxy
for silicon is confirmed in the equivalent width maps of Cas A presented in
\citet{Hwang00}. Using these X-ray and optical data sets, we discuss
below our findings regarding large and small scale optical/X-ray comparisons,
possible time delays between the appearance of optical and X-ray emissions, and
selected regions that show no corresponding optical or X-ray emission.

\subsection{Large Scale X-ray and Optical Emission Comparisons}

\begin{figure}
\epsscale{0.75}
\includegraphics{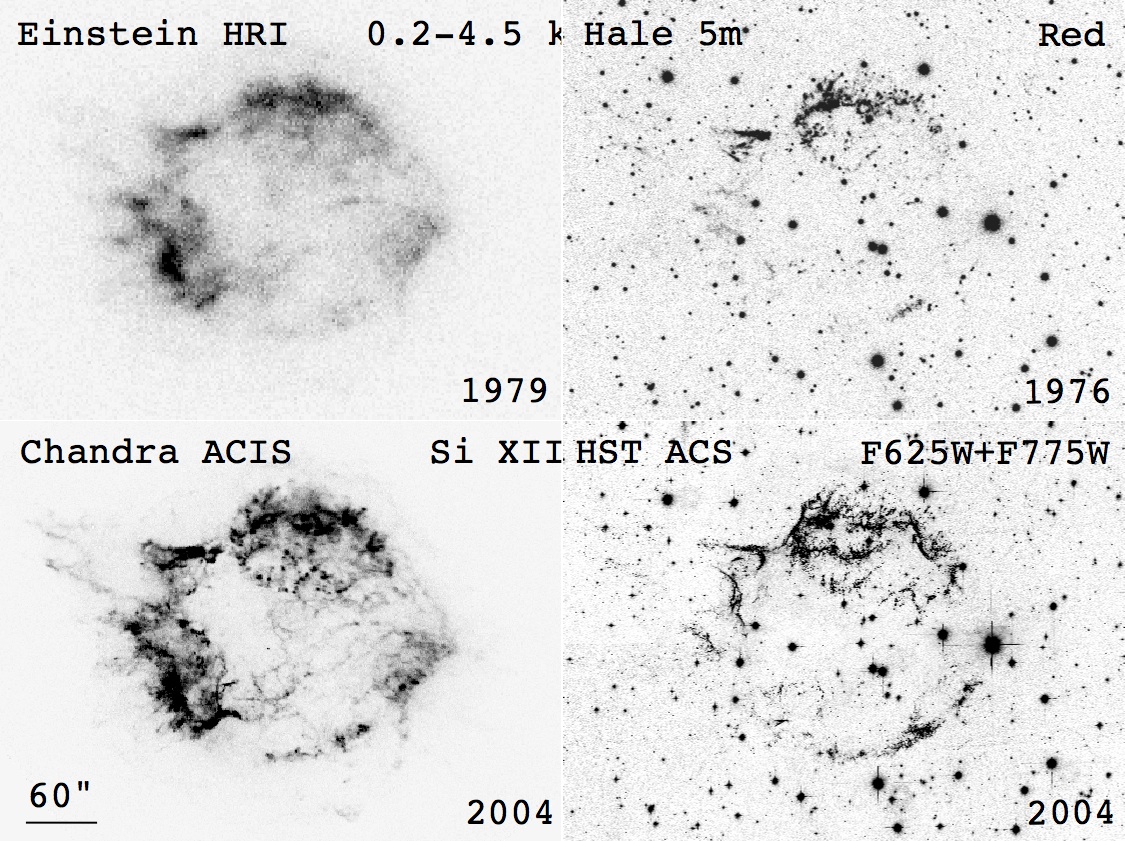}
\caption{Comparison of gross emission differences of Cas A as seen in X-rays and optical. 
Left hand panels show the 1979 {\it Einstein} X-ray image and
a 2004 {\it Chandra} continuum subtracted \ion{Si}{13}  X-ray image.
Right hand panels show broadband red optical images taken at similar epochs to the X-ray images. The optical
images are sensitive to line emission from 
[\ion{S}{2}] $\lambda\lambda$6716,6731, [\ion{O}{1}] $\lambda\lambda$6300,6364
and [\ion{O}{2}] $\lambda\lambda$7320,7330.
The Palomar Hale 5m image is the July 1976 plate PH7252vB while the HST image is a
co-add of March 2004 ACS F625W and F775W images. }
\label{xray_opt_comp}
\end{figure}

From the first high-resolution X-ray image of Cas A taken with the {\sl
Einstein} satellite in 1979, it was clear that Cas A's X-ray and optical
morphologies exhibited significant large scale differences. 
This can be seen in
Figure~\ref{xray_opt_comp}, where we contrast the remnant's X-ray and optical
emissions in the late 1970s and some 25 years later in 2004. 

In the upper two panels of Figure~\ref{xray_opt_comp}, we show the 1979 {\it
Einstein} HRI broadband image of Cas~A along side of a deep 1976 Palomar Hale
5m red optical image.  While there is some correlation between certain emission
regions seen in the 1976 optical and 1979 X-ray images especially along the
remnant's northern limb, there is a poor large-scale optical/X-ray
emission correlation over much of the remnant, especially notable in the
remnant's south, east, and west quadrants.  Most obvious is the lack of bright
optical emission in the 1976 Palomar image coincident with the X-ray bright
eastern and southeastern limbs, and the virtual absence of any optical emission
along the remnant's western limb, where moderately bright X-ray emission is present.

\begin{figure}
\epsscale{0.95}
\includegraphics{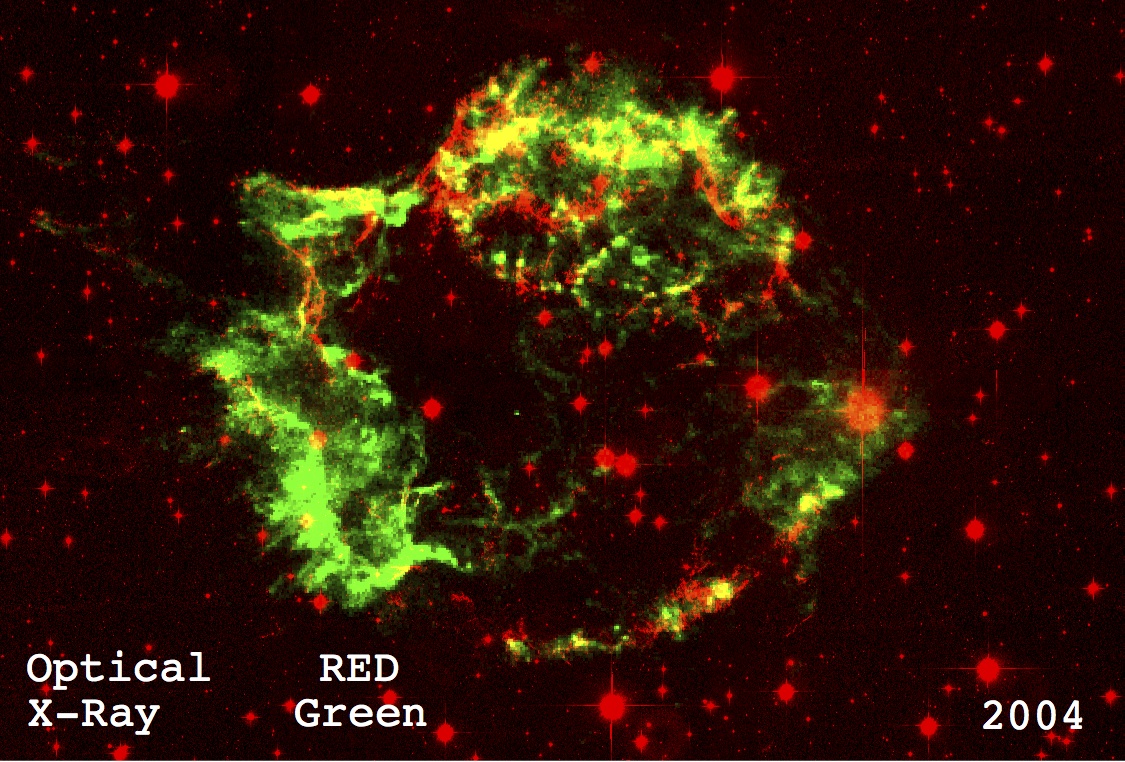}
\caption{A color composite of 2004 images of Cas A. The optical image (red) is
a March 2004 {\sl HST} ACS WFC broadband red image (F625W+F775W) while 
the X-ray image is
a 2004 {\it Chandra} continuum subtracted \ion{Si}{13} image.}
\label{fig:comp_color}
\end{figure}

There is a better correlation between X-ray and optical emissions in 2004
{\it Chandra} and {\it Hubble Space Telescope} images.  In the lower panels of
Figure~\ref{xray_opt_comp}, we show a 2004 continuum subtracted \ion{Si}{13}
X-ray image of Cas A taken with {\it Chandra} ACIS-S on the left and a 2004
{\it Hubble Space Telescope} ACS image on the right where we have co-added the
F625W and F775W filter images to simulate a broad passband band R filter image
approximating the red 1976 Palomar image. While there is now optical emission
along the remnant's southwestern limb which matches the X-ray, there
is still little optical emission in the southeast where the remnant's X-ray
emission is especially strong and extensive. These similarities and differences
can be more readily visible in Figure~\ref{fig:comp_color} where we present a
color composite of 2004 optical (red) and X-ray (green) emissions. 

\begin{figure}
\includegraphics{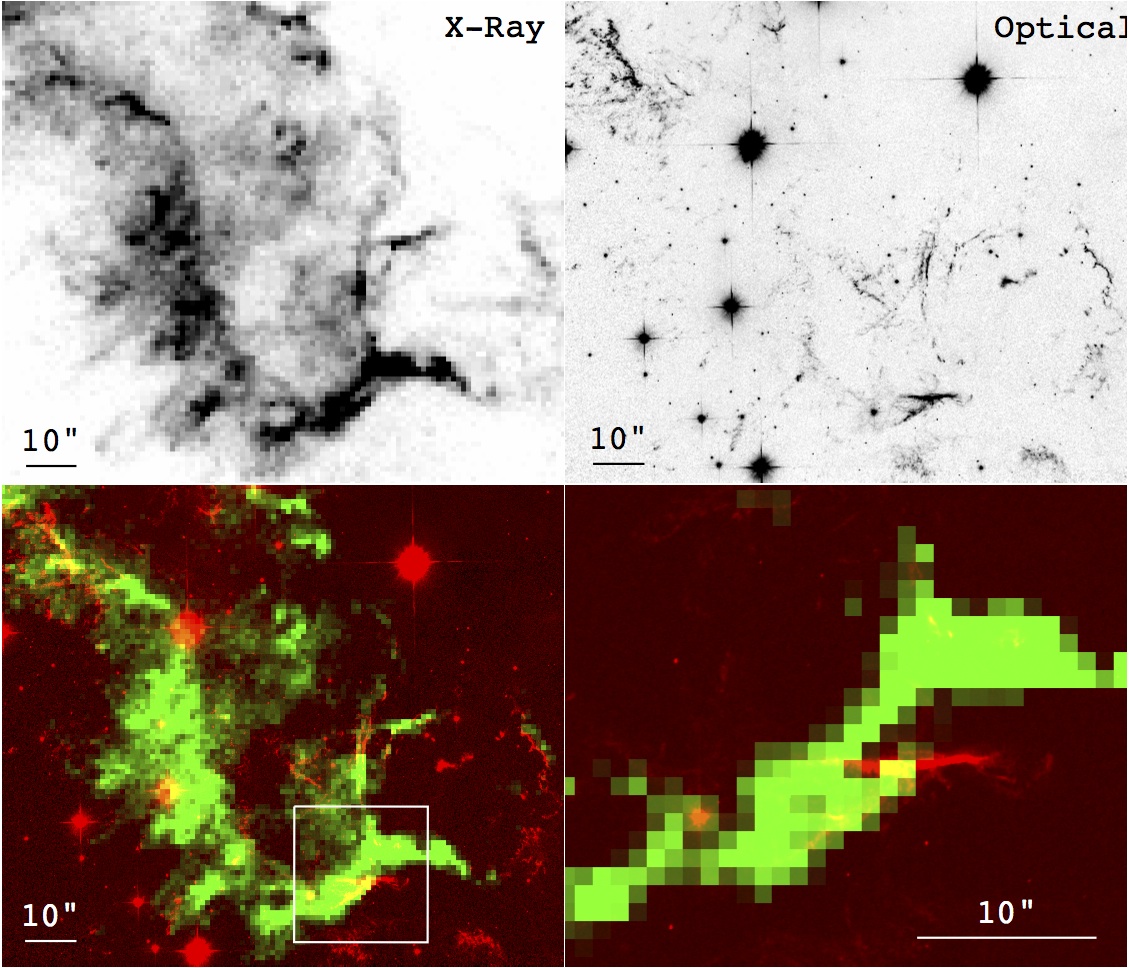}
\caption{Comparison of the southeastern region of Cas A in X-rays and optical.
The upper panels show the same 2004 {\sl Chandra} Si and 
{\sl HST} F625W + F775W images as
in Figure~\ref{fig:comp_color} but now enlarged and covering just the 
remnant's southeastern region.
The lower left panel show a color composite of these images 
(optical = red, X-ray = green).
The white boxed region in the lower left panel marks where the 
X-ray and optical emission is brightest
and shown enlarged in the lower right panel. }
\label{Southeast}
\end{figure}

The southeastern region of Cas~A has consistently shown especially poor
X-ray/optical emission correspondence.  In order to examine this region more
closely, we show in Figure~\ref{Southeast} an enlarged comparison of 2004 X-ray
and optical emissions along the remnant's southeastern limb (upper panels)
along with a color composite of these X-ray (green) and optical (red) images
shown in the lower panels. Features with coincident X-ray and optical emissions
appear yellow.  As can be seen from this figure, one finds only sparse spatial
correlations between optical and X-ray emission features in this portion of the
remnant.  In fact, the brightest X-ray feature in this region, marked by the
white boxed region and shown enlarged in the lower right panel of
Fig.~\ref{Southeast}, lies substantially offset from bright neighboring optical
features.  In $\S$4 below we will discuss possible explanations for such
anti-correlations.

\begin{figure}
\includegraphics{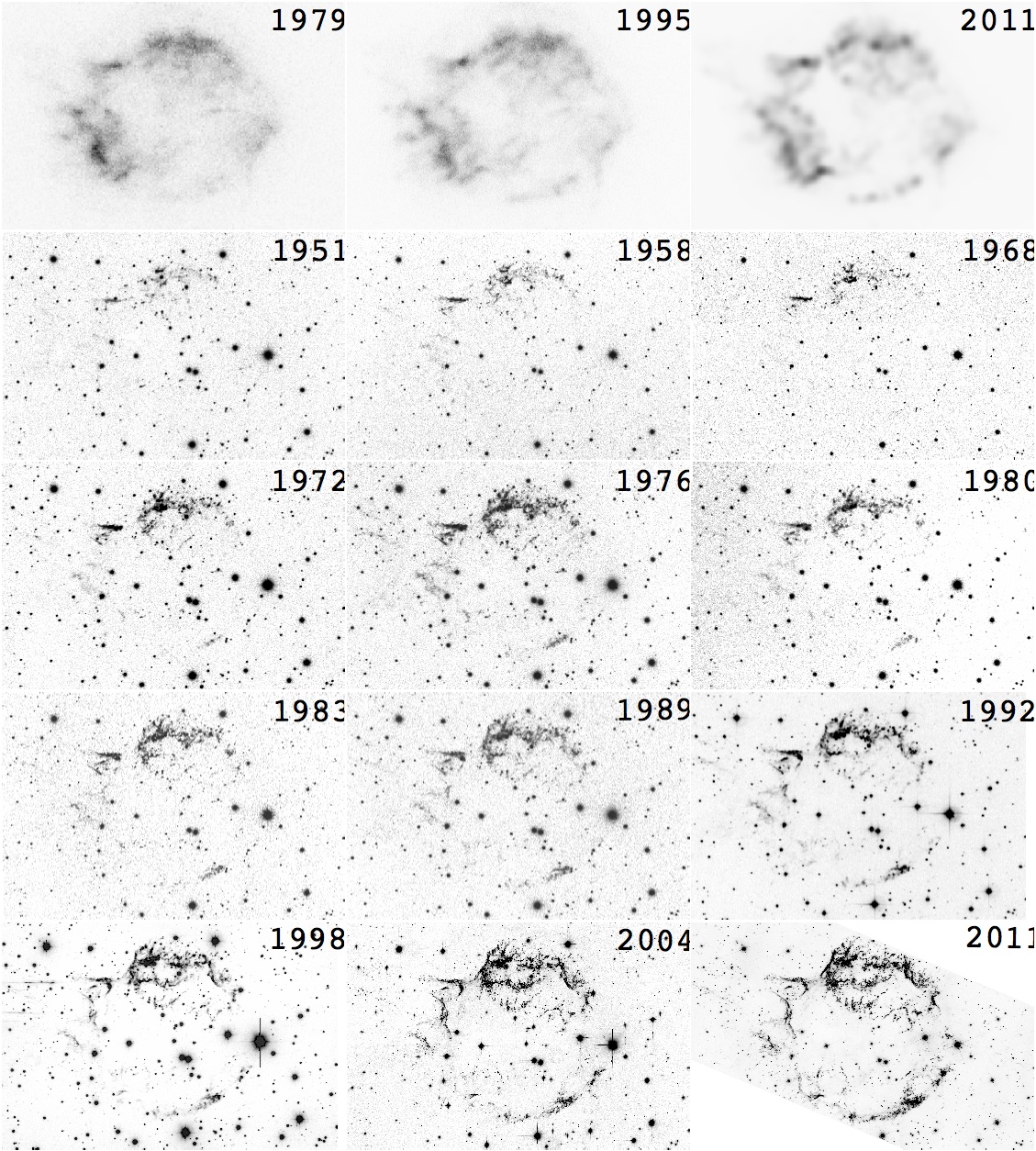}
\caption{Comparative emission changes of
Cas A's X-ray and red optical emission over 31 and 60 years, respectively.{\it
Upper Panels}: {\sl Einstein}, ROSAT, and {\it Chandra} ACIS-S images of Cas A
showing an apparent increase of clumpy emission features from 1979 to 2011.
The Chandra image has been smoothed by a 9 pixel Gaussian to approximate the
resolutions of the {\sl Einstein} and ROSAT images.  {\it Lower 4 rows}: Cas A
in optical broadband red emission images from 1951 to 2011. The 1951 -- 1989
images are Palomar Hale 5m plates PH563B, PH3033S, PH5254vB, PH6249vB,
PH7252vB, PH7766vB, PH8192vB, PH8206vB, the 1992 and 1998 images are MDM 1.3m
and 2.4m images, while the 2004 and 2011 are {\sl HST} ACS F625W + F775W and
WFC3 F098M images (see Table 1 for details). Note the considerable brightening
of the remnant's optical emission along northern and southern limbs, most
dramatically seen between the early 1950s and the 1970s, but continuing up to
the present.}
\label{fig:60yrs}
\end{figure}

It is important to note that unlike much of its X-ray emission, the remnant's
overall optical emission has shown substantial changes over the last few
decades, becoming both brighter and more pervasive.  This is illustrated in
Figure~\ref{fig:60yrs} where we present X-ray and optical images of the remnant
over time spans of 31 and 60 years, respectively.  In the top row we show the
1979 {\sl Einstein} image, the 1995 ROSAT image along with a November 2010
(listed here as 2011) \ion{Si}{13} {\it Chandra} X-ray image of Cas A smoothed
to match the angular resolution of the earlier images.  In the 12 lower panels,
we show optical red passband images of Cas~A from 1951 to 2011.  This figure
shows that while Cas A's overall X-ray emission has undergone only fairly minor
large scale changes over the last three decades, its optical appearance has
gone from one of faint and sparse emission structures, to a bright and
extensive emission morphology in 60 years. 

Although the optical observations shown in Figure~\ref{fig:60yrs} cover a time
span that is roughly twice that of the X-ray images, it is apparent that the
bulk of the remnant's gross X-ray emission evolves on a slower timescale
than its optical emission. That is to say, Cas A looks fairly similar in 2011
as its did in 1979 when viewed in X-rays, whereas the remnant's optical
appearance has changed substantially in many regions on timescales of a decade 
or less.

\subsection{Small Scale X-ray/Optical Emission Correlations}
\label{sub:smallscale}

At {\sl Chandra's} ACIS resolution, \citet{laming03} found a significant
fraction of Cas A's X-ray emission consists of knots or small-scale clumps.
They noted that most of these clumpy X-ray features do not coincide with the
remnant's high-velocity optical ejecta knots and concluded that they represent
only mildly over-dense regions rather than true, high density ejecta knots.

However, there are a number of small-scale features where optical and X-ray
emissions do appear to positionally coincide, especially in cases of recently
brightened optical and X-ray ejecta clumps. Below, we present and discuss a few
of such cases where it appears that the remnant's advancing reverse shock has
resulted in correlated optical and X-ray changes and we explore both the
spatial and temporal properties of such features.

\begin{figure}
\includegraphics{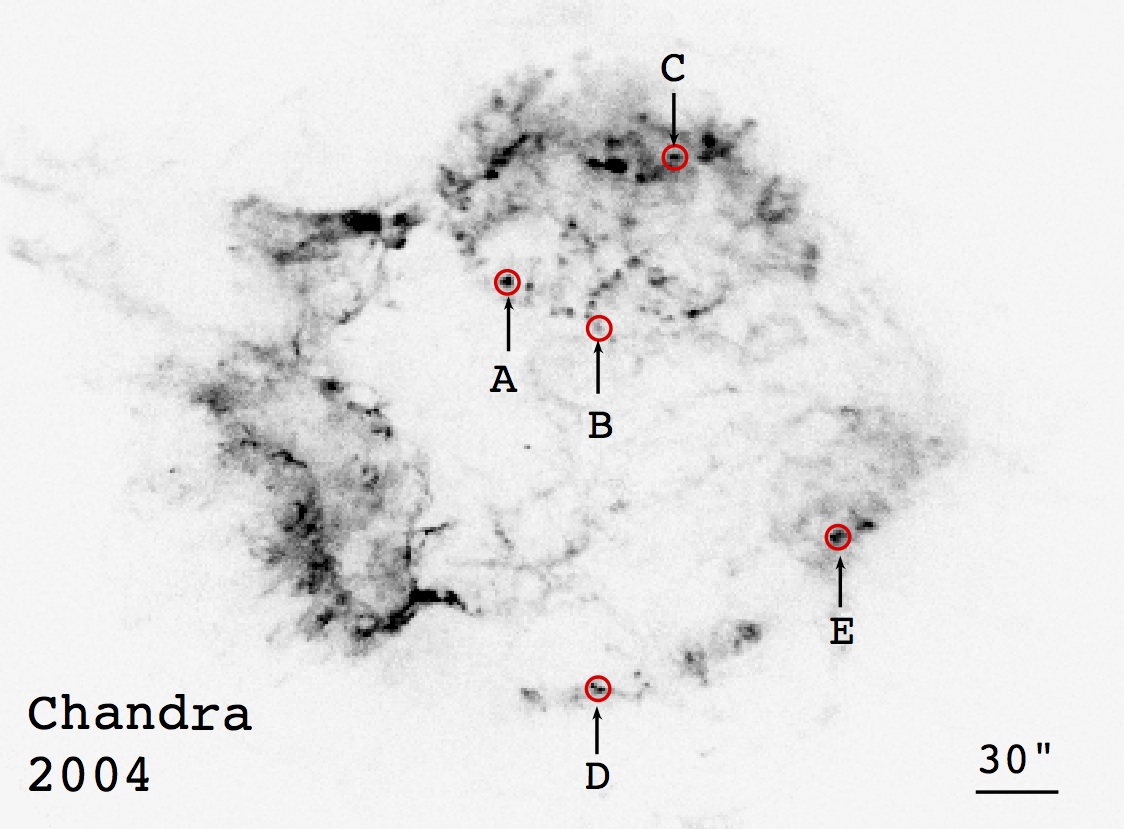}
\caption{The 2004 {\it Chandra} ACIS-S image of Cas A with five regions
marked which showed significant brightening.}
\label{fig:comp_color_xray}
\end{figure}

Figure~\ref{fig:comp_color_xray} shows the locations of five small-scale
features which have exhibited significant brightening in both X-rays and in the
optical within the last few decades.  These features, marked as A--E on the
2004 {\sl Chandra} image, appear to be either absent or weak on earlier
Einstein or ROSAT images yet currently rank among the remnant's brightest
small-scale X-ray features.  The X-ray flux evolution of these five
regions are listed in Table 2, shown in Figures 7--11, and briefly
discussed below.

\begin{figure}
\includegraphics{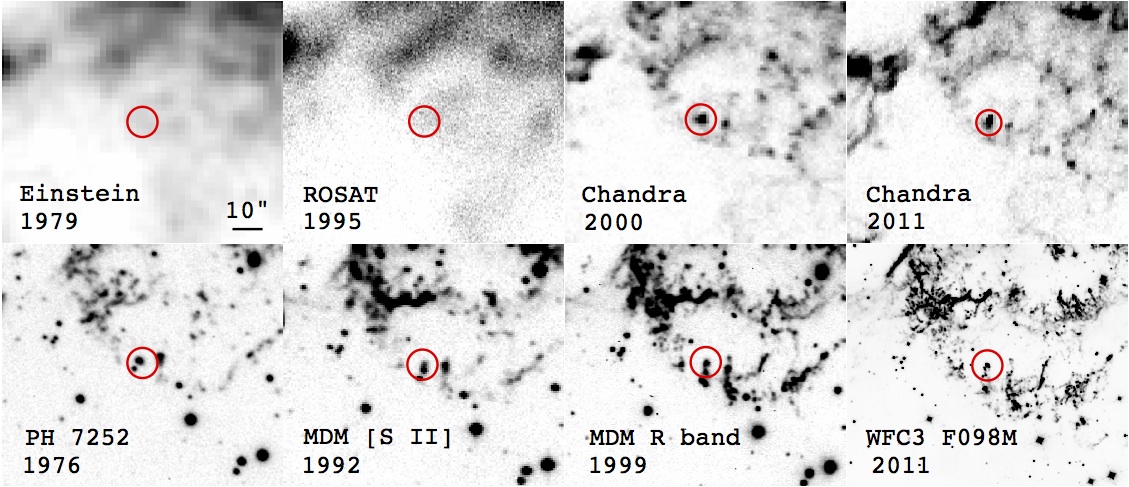}
\caption{ Comparison of X-ray and optical emission for Feature `A'.
{\it Upper panels}: Einstein, ROSAT and Chandra X-ray images with a
circle 10$\arcsec$ in diameter centered on the Feature A's 
approximate location.
{\it Lower panels}: Images of coincident optical emission for this feature.
Note: The circles are centered on the emission feature and 
follow the feature's proper motion. In the 1976 and 1992 images, the optical
knot is seen to cross in front of or behind a brighter 
stationary QSF emission knot.  }
\label{feature_A}
\end{figure}

Feature A: The upper row of panels in Figure~\ref{feature_A} show 1979, 1995,
2000, and 2011 X-ray images of a region, marked by $10\arcsec$ diameter red
circles in the remnant's north-central regions which brightened in X-rays
between 1995 and 2000 (see Table 2). Similar circles in the lower panels mark
the coincident optical ejecta knot seen in images taken between 1976 and 2011.
Whereas by 2000 this feature was one of Cas~A's brightest X-ray emission knots,
the coincident optical emission is weak and unremarkable in its optical
emission from 1992 up to 2011. While virtually absent in the Palomar image
prior to 1972 and greatly complicated by the presence of a nearby and
stationary QSF in the earliest images, the knot had an estimated R2 magnitude
of 20.9 in 1976 but 20.1 and 19.4 in 1996 and 1999
respectively, in line with its brightening in X-rays.  

\begin{figure}
\includegraphics{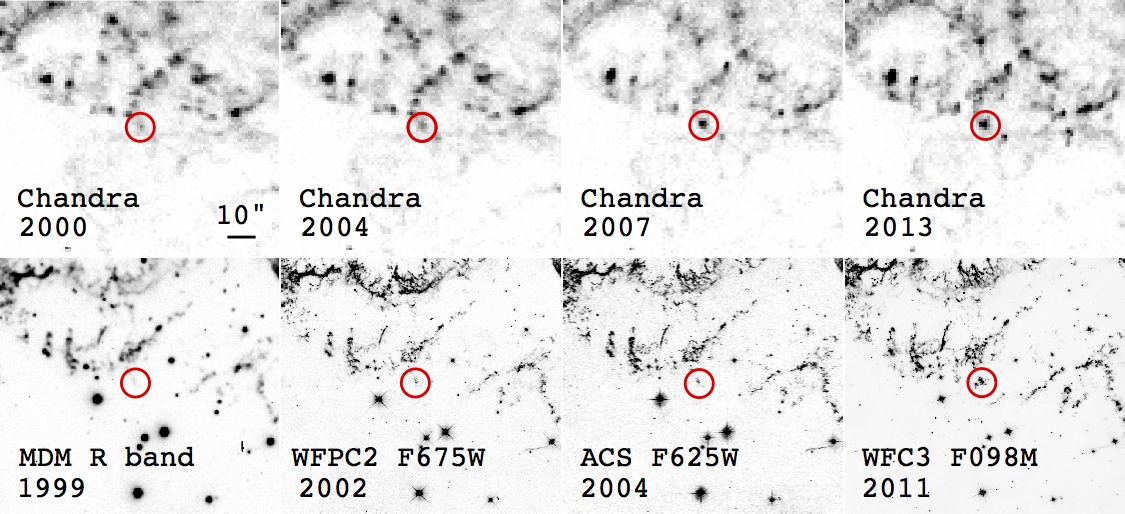}
\caption{Comparison of X-ray and optical emission for Feature `B'.
{\it Upper panels}: 2000 to  2013  {\sl Chandra} X-ray images of Feature B.
The red circles (dia. = 10$\arcsec$)
are centered on Feature `B'. 
{\it Lower panels}: Optical images of the area around 
Feature 'B' at the same scale as the X-ray images.  
The ejecta knot responsible for the X-ray emtting Feature `B'
brightened considerably between 1999 and $2002$ in the optical and between 2000
and $2004$ in X-rays.  The optical image sections shown are approximately
centered on the emission feature but do not follow the feature's proper
motion. }
\label{feature_B}
\end{figure}

Feature B: Considerable clumpy emission is seen in the 2004 {\sl Chandra} image
of the remnant's north-central region (Fig.~\ref{fig:comp_color_xray}).
Enlargements of four {\sl Chandra} X-ray images taken between 2000 and 2013
centered on one of these emission knots labeled Feature B are shown in the
upper panels of Figure~\ref{feature_B}, along with corresponding optical images
covering the epochs 1999 to 2011.  Like that seen for Feature A, this region 
exhibited a significant X-ray and optical brightening after 2004 and will be
discussed further in $\S$3.3.1 below. 

\begin{figure}
\includegraphics{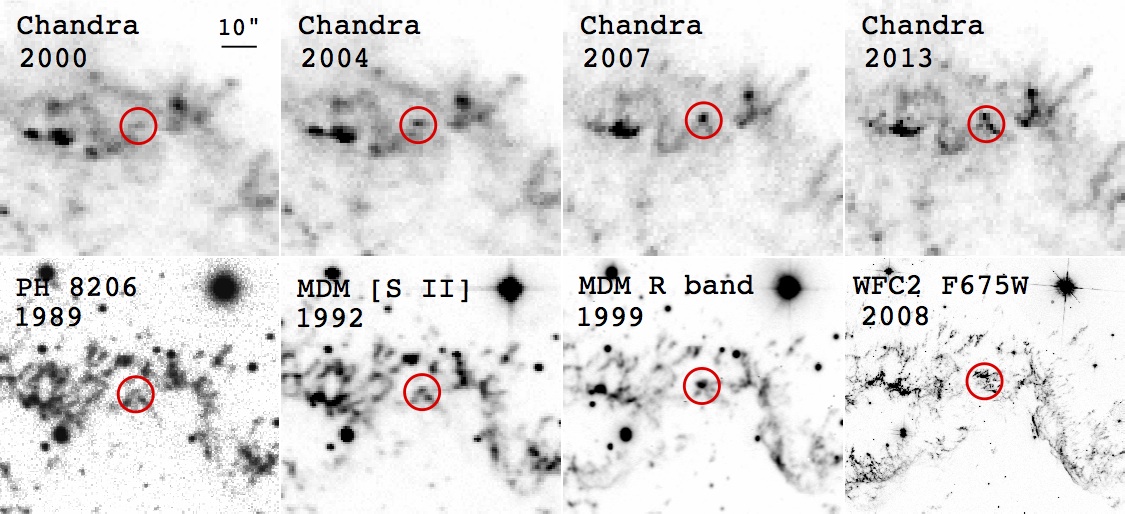}
\caption{ Comparison of X-ray and optical emission for Feature `C'.
{\it Upper panels}: X-ray images of emission Feature C.
The red circles have a diameter of 10$\arcsec$ on the X-ray images.
{\it Lower panels}: Optical images. The red circle shown here has a 
diameter of 10$\arcsec$.
The optical images are: Palomar 5m images PH3033S, PH5107, PH6249vB, PH7252vB,
PH8192vB, PH8206vB, an MDM 1.3m image taken using a broad [\ion{S}{2}] 
interference filter,
and an MDM 2.4m image taken with an R filter.
The optical image sections shown are approximately centered
on the emission feature and follow the feature's proper motion northward.
These images show a dramatic increase in X-ray brightness between 2000 and 2011
and a sharp increase in optical brightness between 1989 and 1992. }
\label{feature_C}
\end{figure}

Feature C: Figure~\ref{feature_C} shows a comparison of an X-ray and optical
emission for the extended Feature `C' located along the remnant's northernmost
rim.  These images reveal a dramatic increase in X-ray brightness between 2000
and 2013 preceded by a sharp increase in coincident optical emission between
1992 and 1999.  The feature's X-ray emission was relatively weak in 2000 but by
2011 was one of Cas~A's brightest emission knots. However, unlike the two
previous cases, its coincident optical emission is relatively bright although
not ranking among the remnant's brightest optical features.  Also, the
evolution of the X-ray emission in this feature appears to follow that seen in
the optical but delayed by several years. For example, the feature's X-ray
emission was only apparent in 2000 whereas optically it became detectable by
the late 1980s, showed marked increases in 1992 and 1999 mirrored later in the
X-rays between 2004 and 2007.

\begin{figure}
\includegraphics{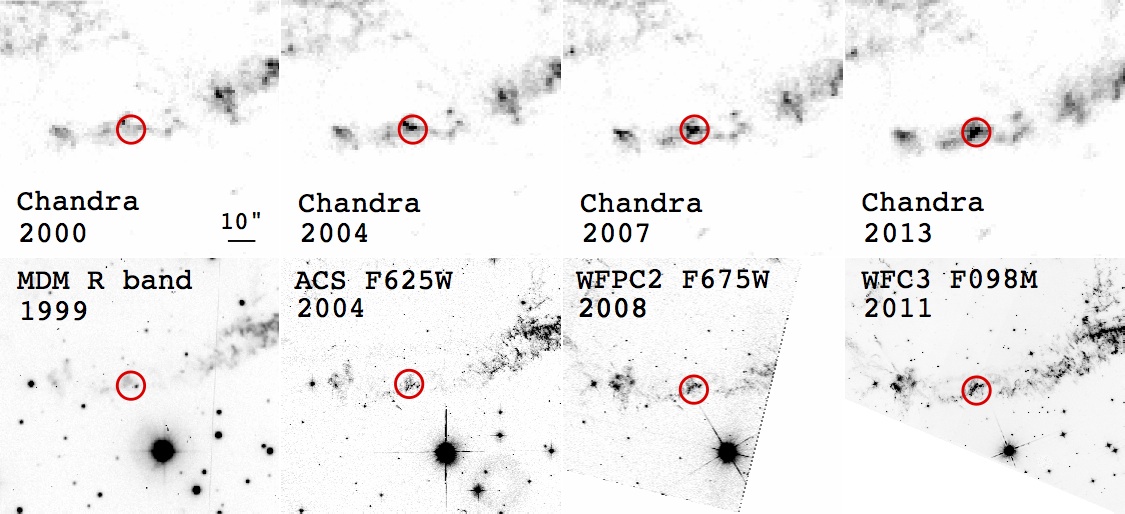}
\caption{ Comparison of X-ray and optical emission for Feature `D'.
{\it Upper panels}: X-ray images of emission Feature 'D'.
The red circles have a diameter of 10$\arcsec$.
{\it Lower panels}: Optical images of this feature. 
The images are: Palomar 5m images PH8202vB,
a 1992 MDM 1.3m image taken using a broad [\ion{S}{2}] interference filter,
and an MDM 2.4m image taken with an R filter, and a 2008 {\sl HST} 
WFPC2 F675W image.
The optical image sections shown are approximately centered
on the emission feature and follow the feature's proper motion northward.
These images show a steady increase in X-ray brightness between 2000 and 2011
and a sharp increase in optical brightness between 1992 and 1999. }
\label{feature_D}
\end{figure}

Feature D: This southern limb X-ray emission clump, shown in
Figure~\ref{feature_D} consists of a small complex of knots which became bright
in both X-rays and optically after 2000.  By 2004, this feature was one of the
brightest X-ray knot complexes along the remnant's southern limb.  The
feature's morphology is somewhat extended and showed significant changes
between 2004 and 2007.  

Feature E: Located along Cas A's western limb, this feature exhibited a
dramatic increase in optical flux after 1996. Its X-ray
flux also increased after 2000, showing a second sharp increase in
X-ray flux after 2010. There is some indication of a time delay of around
three to five years between the onsets of optical and X-ray brightening.
However, this is hard to confirm, because like Feature D, this is a complex of
emission knots and clumps with similar but not exactly matching optical and
X-ray morphologies. 

The optical morphology possibly tying these five X-ray/optical co-spatial
emission regions together is examined in Figure~\ref{fig:region_a2e}. As shown
in these March 2004  {\sl HST} images, all five small-scale regions (A through E)
exhibit a structure consisting of many small bright knots plus 
considerable surrounding diffuse emission.  A similar small-scale structure is
also seen for the X-ray bright northeastern feature known as ``Filament 1''
\citep{BM54} and noted by \citet{fabian80} as a case of especially good
X-ray/optical co-spatial emission.  The combination of both a highly
inhomogeneous density structure and the recent passage of the remnant's reverse
shock may account for the co-spatial X-ray and optical emission correlations
observed for these regions.

\subsection{Optical and X-ray Positional and Temporal Offsets}

In addition to co-spatial X-ray and optical emission, we also found positional
and temporal offsets between Cas A's X-ray and optical emission.
Figure~\ref{fig:comp_color_NE} shows an enlargement of the color X-ray and
optical composite shown in Figure 3 but now covering just the north-central
portion of the remnant.  This region was selected because it contains several
coincident small scale optical/X-ray knots.  Two emission features of special
interest are marked by white circles.  Feature 1 is an example of a positional
offset between associated clumpy X-ray and optical emissions, while Feature 2
is a case of relatively bright X-ray emission patch having no corresponding
present optical emission but coincident with an optical feature visible decades
prior but no longer easily detected. 

\subsubsection{X-ray/Optical Emission Offsets}

We begin by presenting evidence for small-scale spatial offsets between a
clump's X-ray and optical emission.  The small knot marked as Feature 1 is one
such example. It consists of a small ($< 5\arcsec$) optically bright knot
similar to several other knots located to the north and northwest of it and
seemingly forming a chain of optically bright knots each with some coincident
or nearly coincident X-ray emission.  As can be seen in this figure, the
optical knot marked as Feature 1 exhibits some associated X-ray emission but
which is offset slightly to the west. 

This positional offset is examined in greater detail
in Figure~\ref{fig:Region1} where we show 2000 through 2010 X-ray and
optical images along side of a color X-ray/optical composite.  The knot's
optical and X-ray emission fluxes increase substantially over this period.  At
all four epochs between 2000 and 2010, the knot's associated X-ray emission is
not precisely coincident with the knot's optical emission, instead offset by a
few arcseconds to the west.  This is much greater than possible
alignment WCS errors between the optical and X-ray images. 

A clue to the nature of such X-ray/optical emission offsets may come from the 
positional offset case seen in a line of apparently recently shocked ejecta
located near the remnant's south central region, shown in
Figure~\ref{fig:Parenthesis}.  Comprising part of small arcs of filaments and
referred to as the `parentheses' by \citet{delaney10} (see their Fig.\ 13).
This emission arc lies on the facing (blueshifted) hemisphere of Cas~A
\citep{MF2013} and has been interpreted as the walls of a reverse shocked
ejecta cavity \citep{MF2014}.

The morphology of this filamentary arc suggests the remnant's reverse shock has
passed over a line of dense ejecta clumps, leading to Rayleigh-Taylor head-tail
structures along with some trailing diffuse emission seen visible in the {\sl
HST} ACS 2004 and WFC3 2011 images. Associated X-ray emission can be seen offset and
west (i.e., behind) the line of optical knots again by a few arc seconds
($\simeq 1 \times 10^{17}$ cm).  The positional offset behind this arc of
filaments along with the trailing diffuse emission is present from 2000 through
2011 and suggests X-ray emission is enhanced in lower density regions
associated with and derived from post-shocked higher density ejecta clumps.
This interpretation is supported by spectroscopic data indicating substantial
velocity shears ($\simeq 1000$ km s$^{-1}$; \citealt{Fesenetal2001}) in some
parts of the remnant.

\subsubsection{X-ray/Optical Emission Delays}
\label{sub:temp_delays}

Some compact X-ray emission features seen in the {\sl Chandra} images have
little or no coincident \citep{laming03}. One example is Feature 2 marked in
Figure~\ref{fig:comp_color_NE}.  As shown in the mosaic of optical images
covering the time span between 1972 and 1996 presented in
Figure~\ref{fig:Region2}, a relatively faint patch of optical emission appeared
in the early 1970s, brightened in the 1970s and early 1980s but then faded
considerably, becoming barely detectable by the mid-1990s. Summing the flux in
a $5'' \times 5''$ aperture box centered on this feature, we estimate R2 
magnitudes to be 23.3 in 1970, 21.1 in 1972, 20.1 in 1976, 19.5 in 1983, 21.7
in 1989, and 21.5 in 1992.  Deep {\sl HST} images taken in 2000 and later
reveal very little in the way of optical emission in this region.  In contrast,
{\sl Chandra} images of this region show a relatively bright patch of X-ray
emission which has remained roughly constant in strength from 2000 through
2013. The nature of such X-ray bright but optically faint ejecta may be related
to low density ejecta regions which initially give rise to some faint optical
emission but relatively strong X-ray flux at late times.

A similar optical/X-ray emission evolution can be seen in the northern extent
of a thin chain of emission knots roughly 10 arcsec northeast of Feature 2 (see
Figs.\ \ref{fig:comp_color_NE} and \ref{fig:Region2}). 
In 1972, the northern most knot of this chain appears
fairly bright in 1972 and 1976 Palomar images but has faded considerably by the
1990s, so much so as to be barely detectable in 1992 and 1996 MDM images.
Despite its optical fading, associated X-ray emission is relatively strong as
can be seen in the lower right-hand panel of Figure~\ref{fig:Region2}
showing a color composite of
2000 {\sl Chandra} and {\sl HST} images.   

Opposite cases can also be seen, where one sees strong optical emission but no
associated X-ray emission.  Extensive and ever brightening optical emission
along the remnant's northwestern limb first appearing in the early 1980s
exhibits no obvious corresponding X-ray emission, even weakly. Yet this region
exhibits one of the clearest views of the advancement of the remnant's reverse
shock \citep{Morse2004}.  Figure~\ref{figure_NW_arc} compares the region's
optical emission with that seen in the X-rays.  Whereas the westernmost portion
of the circled region shows both bright optical and X-ray emission, the eastern
portion exactly where the remnant's advancing reverse shock front can be easily
identified, has no obvious X-ray counterpart.  If this is a case of a delay of
associated X-ray emission and not simply a density effect limiting the X-ray
emission, then the X-ray/optical time lag must be greater than 20 years.

Finally, Figure~\ref{fig:EastRegions} shows a case of an X-ray bright filament
which slowly formed between 2000 and 2010 in the northeast near the base of the
NE jet, exhibits only faint optical emission (marked in the figure as ``Feature
3'').  Optical emission in this location is extremely faint prior to 2000 and
only weakly visible in deep {\it HST} WFC3 F098M images taken in 2011. Using
the {\it ciao} tool \texttt{specextract}, we extracted X-ray spectra for this
region and fit the He-like silicon line centroid and measured the 1.75--2.0 keV
line+continuum flux. The results from this analysis are listed in
Table~\ref{tab:xray_rs}.  Over the 12 years for which the X-ray data were
analyzed, the fitted line centroid increases by $\gtrsim$ 10 eV while the
1.75--2.00 keV flux increases by nearly a factor of 3. This may be an example
of a ``young'' and low density region that shows only X-ray emission. The east
and southeast regions of Cas A, which show extensive diffuse X-ray emission but
little or no correlated optical emission (see Fig.~\ref{Southeast}) may be
other examples. A lack of strong optical emission in these X-ray bright regions
may also be related to hydrodynamical effects that can smooth out ejecta
inhomogeneities \citep{wong13}.

\section{Discussion}


From the comparisons presented in $\S$3, we have identified four cases of
X-ray/optical correlations or anti-correlations: 1) good optical/X-ray spatial
and temporal correlations for small scale features which appear to have
recently brightened due to a recent interaction with the remnant's reverse
shock front, 2) features whose optical brightening precedes the onset of
associated X-ray emission, 3) optical emission and X-ray emission peaks which
appear spatially offset but present at the same time, and 4) regions showing no
corresponding X-ray or optical emissions.  Below we explore various physical
situations that might give rise to these cases.

\subsection{Physical Parameters}

We begin our discussion of X-ray/optical correlations by first laying out
the general physical parameters and timescales involved.  We assume that the
supernova ejecta can be modeled as a multiphase plasma with a diffuse component
that contains much if not the bulk of the material and gives rise to the
observed X-ray emission. Additionally, we assume there is a significant population of
compact, dense ejecta ``clumps'' that are mainly responsible for the remnant's
observed optical emission.  There are relevant timescales associated with each
of these components and our aim is to describe a relation between them
that is consistent with X-ray and optical properties described previously.

For ejecta that has not yet been shocked by the remnant's reverse shock front,
both the diffuse and clumpy components will be co-moving.  We model the low
density component by an expanding ejecta cloud with number density of
$n_{\mathrm{ej}}$ $\sim$ 1.0 cm$^{-3}$ \citep{laming03}. This number density of
$\sim$ 1.0 cm$^{-3}$ has recently been confirmed observationally by low
frequency radio absorption measurements where the authors measured a density of
$\approx$ 4 cm$^{-3}$ \citep{delaney14}.

Embedded within the low density ejecta may be small clumps with densities
$n_{\mathrm{c}}$ $\sim$ 100--1000 cm$^{-3}$. In this situation, we assume
that the two components are co-moving and in pressure equilibrium,
otherwise the dense knots would experience mass ablation due to ram-pressure
stripping as they travel through the diffuse component. Thus,

\begin{equation}
n_{\mathrm{ej}}k_B T_{\mathrm{ej}} = n_{\mathrm{c}} k_B T_{\mathrm{c}} \, ,
\end{equation}

\noindent where $k_B$ is Boltzmann's constant and $T_{\mathrm{ej}}$ and
$T_{\mathrm{c}}$ are the temperatures of the diffuse and clumped components.
The temperature of the clumps are thus $\sim$ 10$^{-2}$ -- 10$^{-3}$
$T_{\mathrm{ej}}$.  The density of ejecta clumps embedded within a low density
envelope may not instantaneously change between the diffuse and clumpy
components. We will argue below that such low density material surrounding
dense knots may be responsible for some population of correlated X-ray and
optical structures identified in the images described in $\S$3.

Since the diffuse and clumped components are co-spatial and co-moving, they
will interact with the SNR reverse shock at the same time. We assume a 
$\gamma$ = 5/3 gas meaning that radiative and cosmic-ray losses 
are dynamically
unimportant. Main shell ejecta are expanding with a 
velocity of $u_{\mathrm{ej}}$
= 4000-6000 km s$^{-1}$ \citep{minkowski59,vdb71b,law95,delaney10,MF2013}, 
and the diffuse ejecta downstream from the shock (e.g., the shocked ejecta) 
has a velocity (3/4)$u_{\mathrm{ej}}$.

While we assume that the density between the diffuse and clumped 
components varies relatively smoothly, we can define a density 
contrast between 
the clumps and the diffuse component  to be
$\chi$ $\equiv$ $n_c/n_{ej}$ \citep{klein94}. 
It is the same whether we consider
the shocked or unshocked ejecta since the shock compression is the same
in both the diffuse and clumped components. The $n_c$ specified
here is the maximum number density in the clump, and any calculations
will thus represent an upper or lower limit on the 
computed value. 

In the shocked component of the ejecta, we take the ram pressure equilibrium
between the shocked clumps and shocked ejecta as being 

\begin{equation}
n_{\mathrm{ej,s}} u_{\mathrm{ej,s}}^2 = n_{\mathrm{c,s}} u_{\mathrm{c,s}}^2 \, ,
\end{equation}

\noindent
where the ``s'' subscript denotes a shocked quantity. Given that the density
contrast between the clump and the diffuse component is invariant and
is $\chi$ $=$ 100--1000, and the
velocity of the shocked diffuse component is three quarters that of the 
unshocked
diffuse component ($\approx$ 3750 km s$^{-1}$), the velocity of the 
shock that is driven into the clump is then $u_{\mathrm{c,s}} = \chi^{-1/2} 
u_{\mathrm{ej,s}}$ $\approx$ 100 - 400 km s$^{-1}$. For simplicity, we assume
that $u_{\mathrm{c,s}}$ = 250 km s$^{-1}$. 

\subsection{Emission Timescales}

We are now in a position to compute relevant timescales associated with the
rise in X-ray emission and those associated with the optical emission and
subsequent clump destruction. Since unshocked clumps are comparatively cold and
do not emit strongly in the optical, we will estimate their size based on the
size of the shocked clumps observed with {\sl HST}.  \citet{Fesenetal2001} and
\citet{Fesen11} estimate a radius for optically emitting clumps to be $a_0$
$\sim$ 10$^{15-16}$ cm. 

The clump crossing time, $t_{ic}$, is defined as the time it takes the shock 
in the diffuse component to cross the clump:

\begin{equation}
t_{ic} \equiv \frac{2a_0}{u_{\mathrm{ej,s}}} \, ,
\end{equation}

\noindent
and the time for the clump to be crushed by the clump shock is the 
cloud crushing time:

\begin{equation}
t_{cc} \equiv \frac{\chi^{1/2} a_0}{u_{\mathrm{ej,s}}} \, .
\end{equation}

\begin{figure}
\includegraphics{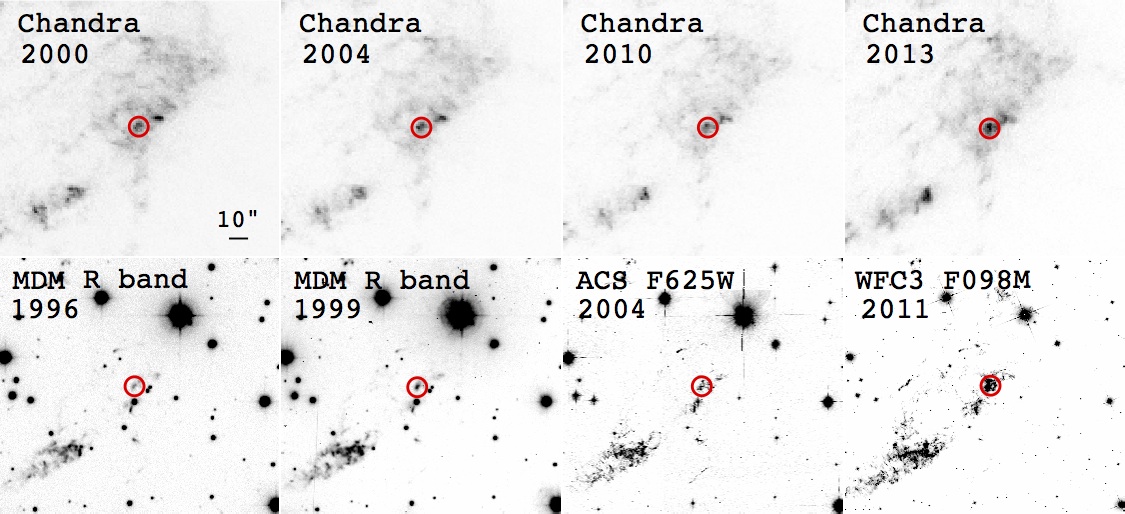}
\caption{ Comparison of X-ray and optical emission for Feature `E'.
{\it Upper panels}: X-ray images of emission Feature E.
The red circles have a diameter of 10$\arcsec$ on the X-ray images. 
{\it Lower Panels}: Optical images of this feature. The images are 
a 1996 and 1999 MDM 2.4m R band image, a 2004 {\sl HST} ACS F625W image, 
and a 2011 {\sl HST} WFC F098M image. The optical images are approximately
centered on the emission features and follow the feature's proper 
motion approximately westward. These images show an increase in optical
flux after 1996, and appeared in X-rays probably prior to 2000.}
\label{feature_E}
\end{figure}

\begin{figure}
\includegraphics{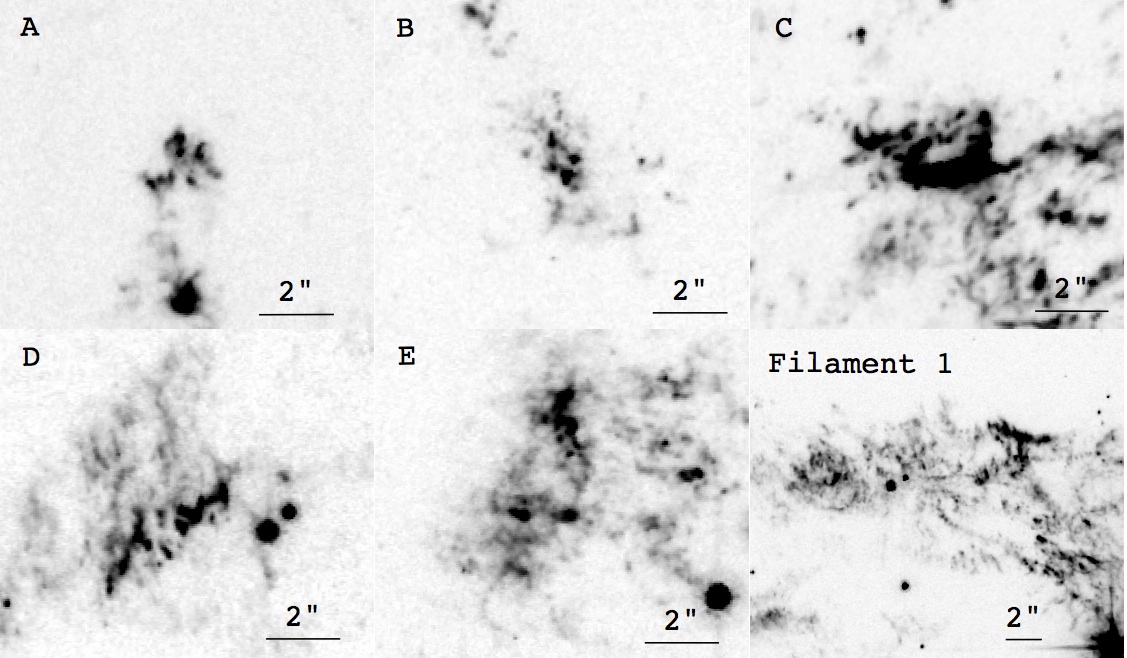}
\caption{Enlargements of the March 2004 ACS F625W+F775W merged 
image showing the detailed
optical structures of Features A thru E plus the bright northeastern
filament known as Baade and Minkowski's Filament 1 which 
is one of the few good X-ray and optical spatial correlations.
In all cases, numerous bright compact optical knots
are seen embedded within or with considerable surrounding diffuse emission.}
\label{fig:region_a2e}
\end{figure}

\begin{figure}
\includegraphics{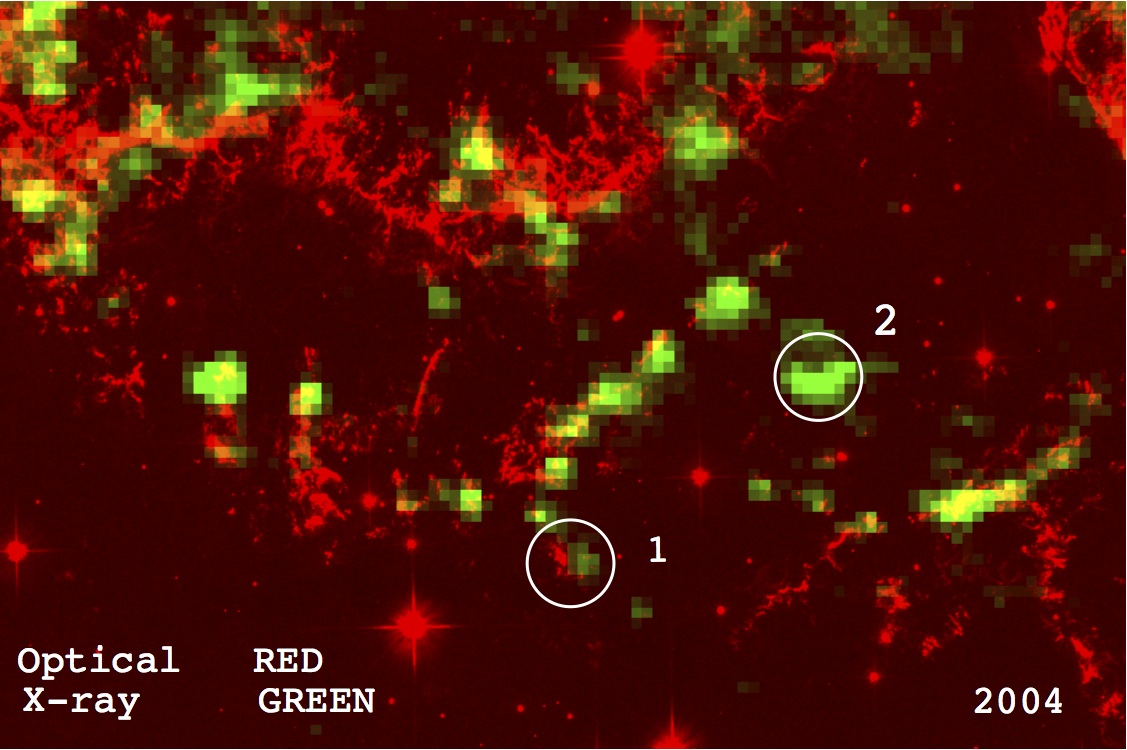}
\caption{Color composite of optical and X-ray images for the north-central
region of Cas A showing several cases of positional correspondance of optical
and X-ray features in 2004.  Two small features are marked by white circles
(dia.\ 8$''$) which are discussed in the text.  Feature 1 is an example of a
positional offset between optical and X-ray emission while Feature 2 marks an
area with significant X-ray flux but lacking appreciable optical emission. }
\label{fig:comp_color_NE}
\end{figure}

\noindent
The resulting crossing time is short, on order of one year. In contrast,
given the size and density contrast of the clump and the velocity of the
surrounding ejecta field, the cloud crushing time is estimated to be
$t_{cc}$ = 5--30 yr. This is consistent with the timescale for 
knots to appear and subsequently disappear \citep{vandenbergh85}, and also
appears consistent with the rise in emission from small scale features like
those seen in Figures~\ref{feature_A} through ~\ref{feature_E}.

While the shock crossing timescale is independent of 
radiative processes, the cooling timescale for ejecta clumps is given by 

\begin{equation}
t_{cool} \equiv \frac{3k_BT_{c,s}/2}{n_{c,s}\Lambda(T_{c,s})} \, ,
\end{equation}

\noindent
where $\Lambda(T_{c,s})$ $\approx$ 10$^{-22}$ erg cm$^{3}$ s$^{-1}$, 
and $T_{c,s}$ is
the temperature of the shocked clump material, T$_{c,s}$ $\approx$ 30,000 K. 
The
cooling time is $\lesssim$ 20 yr, comparable to the timescale for the clump
to be destroyed by hydrodynamical processes. However, the cooling time
does not account for the hydrodynamical destruction and mixing of the clump
into the diffuse component. For comparison, the cooling time for the
X-ray emitting material can also be calculated-- for a temperature $T_{ej,s}$ 
$\approx$ 1 keV, a density $\sim$ 1--10 cm$^{-3}$, and $\Lambda(T_{ej,s})$ 
$\approx$ 10$^{-20}$ erg cm$^{3}$ s$^{-1}$ \citep{hamilton84}, 
the resulting cooling time is $\sim$ 10$^{4-5}$ yr. It has been shown
that conduction can mitigate the effects of radiative cooling in
crushed clouds \citep{orlando05}, but here we note that the heating timescale
(via conduction) is very much longer than the cooling time:

\begin{equation}
\left(\frac{t_{cool}}{t_{cond}}\right) \simeq 10^6 \frac{T^2_{c,s}}{n_{c,s} a_0} < 1 \, ,
\end{equation}

\noindent
so the effects of thermal conduction are not considered.

While the clump is being crushed and eventually destroyed by a 
100 -- 400 km s$^{-1}$
shock driven into it, the associated diffuse component is being 
shock heated. Diffuse 
ejecta is initially cold and partially neutral, with free electrons initially 
supplied
mostly by photoionization of the expanding unshocked ejecta field. For the
diffuse ejecta component, the timescale for thermalization of the electrons
and the timescale to ionize to Li--, He--, and H--like ionization states
are considered. For the diffuse component, cooling is not
important as neither Bremstrahlung emission from the electrons nor line
emission from ions are efficient coolants at the temperatures we are
interested in, as noted earlier.

Electron heating is determined by

\begin{equation}
\frac{\mathrm{d}T_{\mathrm{elec}}}{\mathrm{d}t} = \frac{T_{\mathrm{ion}} - T_{\mathrm{elec}}}{t_\mathrm{eq}} \, ,
\end{equation}

\noindent
where the relaxation timescale $t_{\mathrm{eq}}$ $\propto$ 
$n_{\mathrm{ion}}^{-1}$$\times$T$_{\mathrm{ion}}^{3/2}$. In fact, the
electrons do not need to reach thermal equilibrium with the ions. Instead
they only need to reach temperatures of $\sim$ 1 keV ($\approx$ 1.2$\times$
10$^{7}$ K) to emit at X-ray wavelengths and be detectable through the
intervening interstellar column. The above equation is not 
easily integrated but can be solved numerically.

A second timescale relevant to the X-ray emission is the ionization
timescale, $n_e t$. For X-ray line emission to be important, 
ionization timescales
of $\log{(n_e t)}$ $\simeq$ 9--11 are required. The ionization timescale
is time dependent and also depends on the electron temperature given by

\begin{eqnarray*}
\frac{1}{n_e}\frac{\mathrm{D}f(X^i)}{\mathrm{D}t} & = & C(X^{i-1},T_e)f(X^{i-1}) + \alpha(X^i,T_e)f(X^{i+1}) \\
 & & - [C(X^i,T_e) + \alpha(X^{i-1},T_e)]f(X^i) \ .
\end{eqnarray*}

\noindent
$f(X^i)$ is the fractional ionization state of element $X$ in state
$i$, and $C(X^i,T_e)$ and $\alpha(X^i,T_e)$ are the ionization and 
recombination rates out of and into ion X$^i$, respectively.

We used a one dimensional numerical hydrodynamics code to study the evolution
of the electron temperature and ionization of the diffuse 
component of the ejecta.
This code is based on the numerical code VH-1 developed by \citet{blondin93} 
to which we have added a non-equilibrium ionization calculation and
prescription for electron heating, similar to that described in 
\citet{patnaude09b} although we do not consider cosmic-ray losses here. 

\citet{hwang12} recently completed a census of the X-ray emitting ejecta
in Cas A, and we have used their results for our one-dimensional hydrodynamical
model.
We model Cas A's diffuse ejecta as an $n$ = 10 power law in velocity 
($\rho_{\mathrm{ej}}$ $\propto$ $u_{\mathrm{ej}}^{-10}$) 
expanding into an isotropic
stellar wind ($\rho_{\mathrm{amb}}$ $\propto$ $r^{-2}$) and 
assume 3M$_{\sun}$
of ejecta and an explosion energy of 2$\times$10$^{51}$ erg. For the
progenitor's stellar wind, we assume $v_{\mathrm{wind}}$ = 10 km s$^{-1}$ and 
$\dot{\mathrm{M}}$ = 2$\times$10$^{-5}$ M$_{\sun}$ yr$^{-1}$ 
\citep{chevalier03}. 

The exact choice of explosion and wind parameters affects the final ionization
state and dynamics of the gas but we are mainly looking for a trend in how the
optical emission from dense ejecta clumps relates to the X-ray emission from a
co-moving diffuse component. Consequently, we are not too concerned with the
exact details, only in whether the thermalization timescale for electrons and
the ionization timescale to He-like states is similar to the clump destruction
and radiative cooling timescales. 

In Figure~\ref{fig:model}, we plot the ionization age and electron temperature
of shocked ejecta behind the reverse shock as a function of when the ejecta was
shocked. As seen in this plot, electrons quickly thermalize to X-ray emitting
temperatures, while only after some $\sim$ 20 yr does the ionization parameter
becomes large enough that emission lines from He--like ions become important.
Thus, at least phenomenologically, significant X-ray emission at locations
where optical emission first becomes bright is consistent with an ejecta
structure where dense clumps of ejecta are embedded in a more diffuse medium. 

In the above model, we make no assumptions about the elemental abundance in the
ejecta, and use cosmic abundances. However, since Cas A's
ejecta are highly metal rich, each ionization will, in fact, lead to a higher
electron density per unit time than in the cosmic abundance case. 
Additionally, since the electron heating is $\propto$ Z$^{2}$, the
heating rate in a metal rich plasma will be greater and result in
higher electron temperatures. Thus, the
parameters shown in Figure~\ref{fig:model} 
represent lower limits on the true values.

In the sections that follow below, we describe how the observed X-ray/optical
emission phenomena presented in $\S$3 
can fit within the basic framework described above.

\subsection{Time Delays Between Optical and X-ray Emission}

As discussed in \S\ref{sub:temp_delays}, several recently shocked features
exhibit apparent time delays between the rise of optical and X-ray emission
(e.g., Features C and E).  A particularly interesting example is that of 
Feature 2 in the northern portion of the remnant
(Fig.~\ref{fig:comp_color_NE}). As shown in Figure~\ref{fig:Region2}, a faint
patch of optical emission appeared around 1972, became more easily visible in
1976 and 1983 Palomar images but then gradually faded, becoming nearly
undetectable in deep 1992 and 1996 images. At this location, 2000 and 2004 red
passband {\sl HST} reveal a half dozen small and very faint optical clumps,
some showing some even fainter surrounding diffuse emission.  In contrast, as
shown in the lower right frame of Figure~\ref{fig:Region2}, one sees relatively
strong X-ray emission at this location in the 
{\sl Chandra} image starting in 2000
and in all subsequent images up to the present. 

One can interpret such emission changes within the framework of a cluster of
relatively low density ejecta clumps interacting with the reverse shock some 
20--40
years ago. A shock is driven into a region of moderately
dense ejecta giving rise to detectable but nonetheless relatively weak optical
emission. At the same time, a larger region of lower density ejecta associated
with these optical clumps is eventually heated to $\sim$ 10$^7$ K and ionized
to He-- and H--like charge states. The timescale for this process is some 30
-- 40 years, roughly similar to that seen for Feature 2, and
envisioned in Figure~\ref{fig:ejecta}.  During that time, the
optical clumps cool radiatively, also on a timescale of about 20 -- 30 years,
thus explaining the near disappearance optically by the late 1990s. 

\subsection{Positional Offsets Between Optical and X-ray Emission}

\begin{figure}
\epsscale{0.75}
\includegraphics{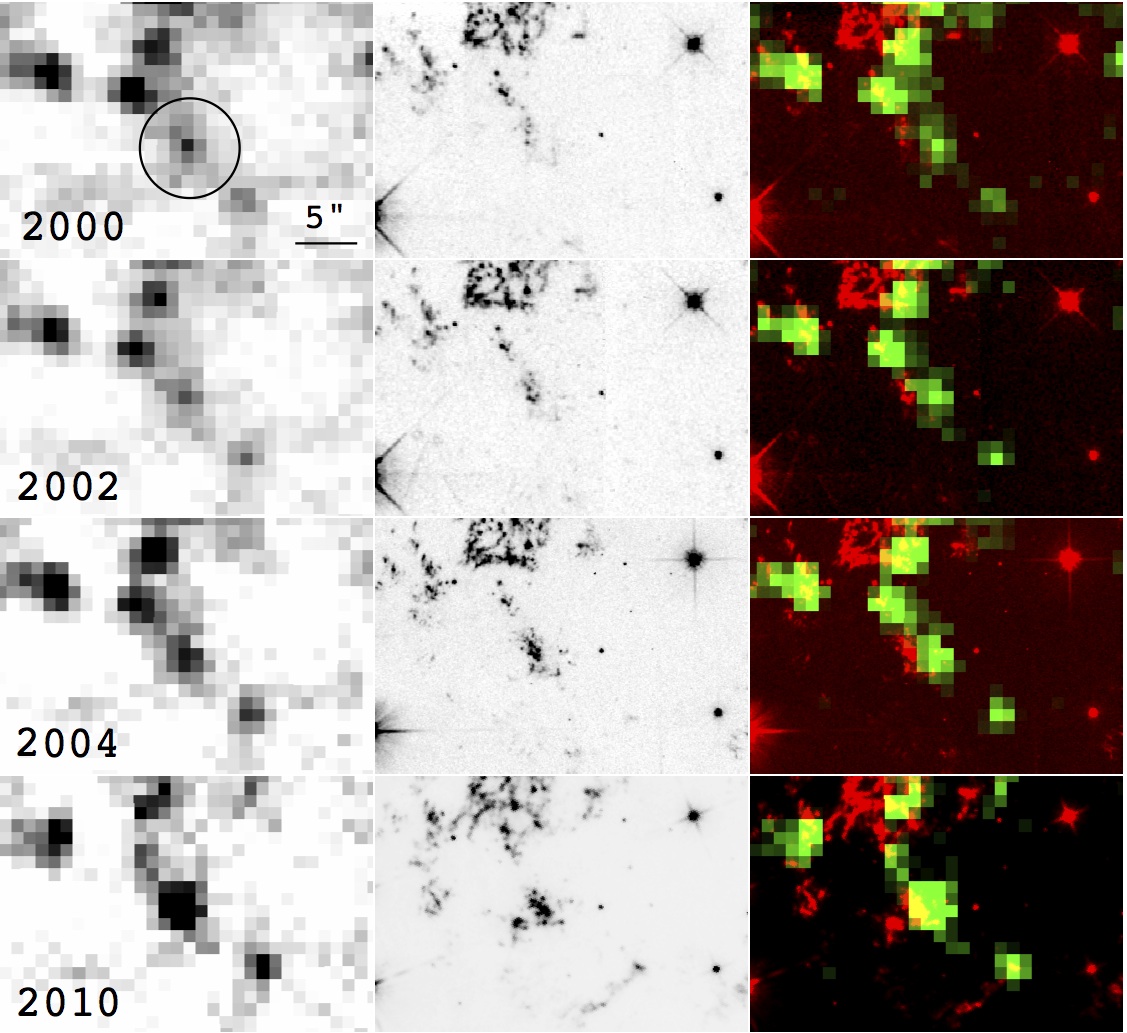}
\caption{A mosaic of 2000 -- 2010 X-ray and optical images of a 
brightening ejecta knot (Feature 1; see Figure~\ref{fig:comp_color_NE}) 
are shown
in Columns 1 and 2 respectively, with a color composite of these 
images shown in Column 3.
The optical images are
{\sl HST} broadband red images while X-ray images (green)
are {\it Chandra} continuum subtracted \ion{Si}{13} image. North is up, 
East to the left.
Note the positional offset to the west of the knot's X-ray emission 
relative to its
optical emission. }
\label{fig:Region1}
\end{figure}

\begin{figure}
\epsscale{0.75}
\includegraphics{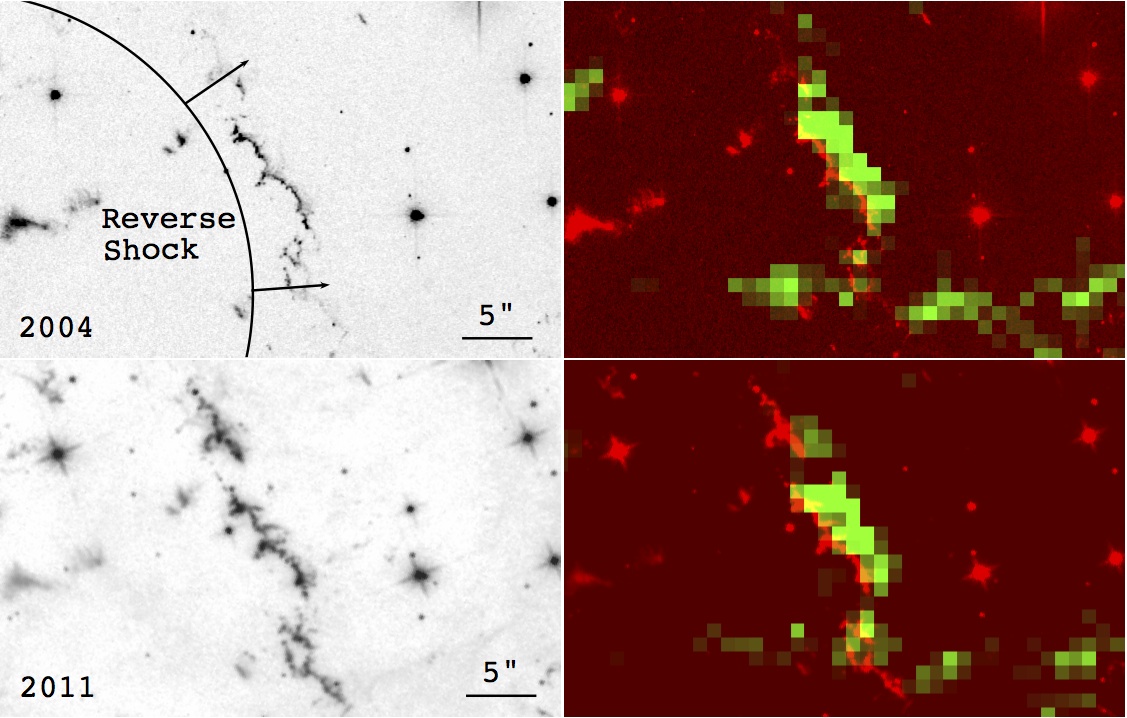}
\caption{A comparison of the March 2004 ACS F625W+F775W and the 2011 WFC3 
F098M images and 2004 and 2011 {\sl Chandra} \ion{Si}{13}
X-ray images of a line of ejecta filaments located southeast of the
center of Cas A referred to as the `Parentheses' by \citet{delaney10}.
Note the positional offset on the X-ray emission behind the line of 
optical emission
where there is apparent head-tail structure in these filaments, most visible 
in the 2011 [S III] {\sl HST} WFC image. The line labeled reverse shock 
does not 
indicate the position of the reverse shock. The
reverse shock lies to the northwest of the optical filaments, near the
position of the X-ray emission.}
\label{fig:Parenthesis}
\end{figure}

\begin{figure}
\includegraphics{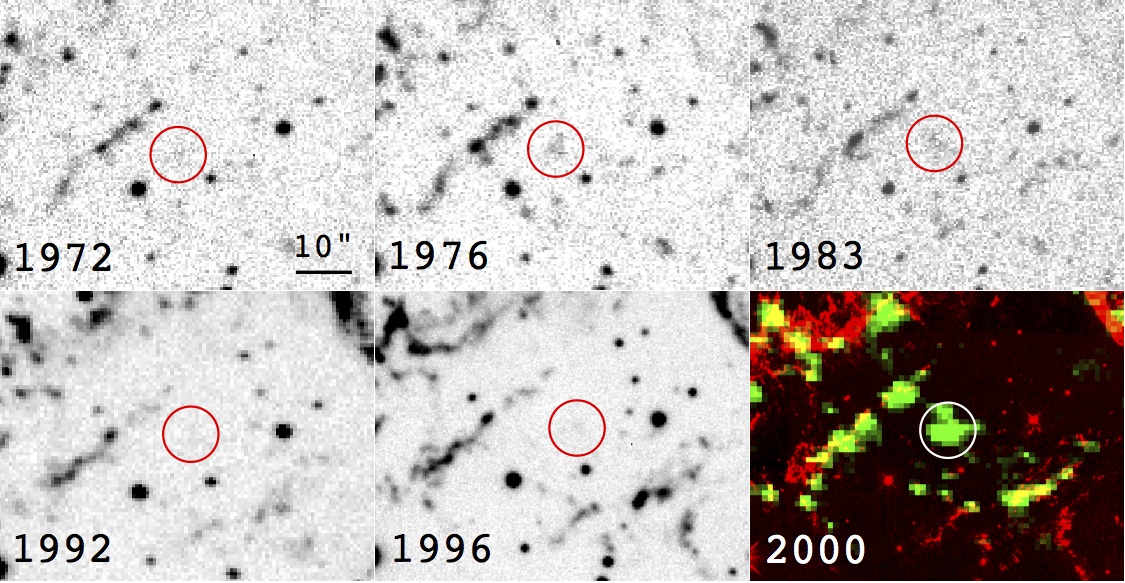}
\caption{A mosaic of optical images of Region 2 marked in 
Figure~\ref{fig:comp_color_NE} covering the time span of
1972 to 1996 along with a color composite of a 2000 {\sl HST} ACS WFC 
broadband red image (F625W+F775W)
and a 2000 {\it Chandra} continuum subtracted \ion{Si}{13} image.
North is up, East to the left.}
\label{fig:Region2}
\end{figure}

There are numerous places in the remnant where this is a clear offset or
separation between a feature's optical and its X-ray emission.
Figure~\ref{fig:Region1} presents one such example. Another is shown in
Figure~\ref{fig:Parenthesis} and it is this line of optical knots that may give
us a clue to what is causing this particular morphology here and elsewhere.

Figure~\ref{fig:Parenthesis} shows a curved line of optical emission along with
displaced associated X-ray emission. These optical knots lie on the facing
hemisphere of the remnant \citep{delaney10,MF2013}.  We have drawn in the
apparent direction of motion in which the reverse shock is traveling for the
line of emission knots.  The morphology of the line of optical emission knots
would suggest that the reverse shock has recently passed over them but their
associated X-ray emission lies to the right (west) and behind the optical
emission and thus closer to the reverse shock's current location than the line
of optical emission knots. 

We propose that such X-ray/optical offsets is, in some cases, generated by shock
induced mass ablation from dense optically emitting ejecta knots.  In
Figure~\ref{fig:opt_tails}, we present a schematic of how this X-ray/optical
offset morphology can be produced. In the first frame, the reverse shock and
the ejecta are seen to be separate. In the second frame, the reverse shock
encounters dense ejecta knots, which are shocked and become bright in optical
emission. Some of the knot material is ablated by the reverse shock, and
advects away from the knots, remaining closer to the reverse shock (third frame
of Fig.~\ref{fig:opt_tails}). This material is of a lower density than in the
knots and so is shock heated to higher temperatures where it can become visible
in X-rays. 

The separation between X-ray and 
optical emission in Figures~\ref{fig:Region1} and~\ref{fig:Parenthesis}
is $\approx$ 2.5$\arcsec$. This corresponds to an advection timescale
of 10 years for the dense knots. It is likely that in these scenarios, the
X-ray emission seen at the reverse shock is not directly associated
with material from the knots that are observed at the same epoch--  
but possibly from knots that previously encountered and were ablated
by the reverse shock. In any event, the low density component at the
reverse shock can be ``refreshed'' by recently ablated material.

A similar set of mass ablation phenomena appears to occur in the NE
``jet'' of Cas A. Here, there is substantial evidence for high velocity ejecta
bullets with space velocities between 8000 and 14,000 km s$^{-1}$
\citep{fesen06}.  As they move away from the remnant's expansion center, they are shock heated via
their interaction with the surrounding circumstellar material. Some mass is
stripped off these ejecta clumps due to Kelvin-Helmholtz instabilities along
the sides of the bullets leading to faint optical emission trails seen in 
{\sl HST} images \citep{Fesen11}. This stripped material will be 
of lower density and consequently shock heated to
high temperatures, and becoming X-ray bright 
\citep{hwang04,laming06} while lagging behind the optically bright
material. In both sets of cases, a positional offset occurs due to mass
ablation of less dense material from the ejecta knots (bullets) which is then
shock heated to higher X-ray emitting temperatures.

\subsection{Regions with Coincident X-ray and Optical Emission}

As described in $\S$3.2, certain regions are
bright in both optical and X-ray emission. These cases may be instances where
optical clumps are embedded within or associated with an extended, lower
density medium like that seen in Figure~\ref{fig:region_a2e}. Both the dense
and less dense regions are shocked at the same time. The optical emission
becomes bright and as it advects away from the reverse shock, the lower density
region becomes bright as well, in X-rays. This is shown schematically in
Figure~\ref{fig:ejecta}. What differentiates this scenario from that presented
in the previous section may be that the maximum density of the optically
emitting material is possibly lower than in the very dense knots, and the
contrast between the dense and diffuse components is less (or the transition
between dense and diffuse components more gradual). It could also occur if 
we are viewing shocked ejecta with significant mass ablation along our line 
of sight. We note that Figure~\ref{fig:ejecta} can qualitatively describe 
both regions that show coincident X-ray and optical emission, as well as 
regions that show a time delay between X-ray and optical emission, if the 
density contrasts and knot envelope structure are appropriately 
chosen.

\subsection{Regions with Neither Temporal or Spatial Correlations}

Just as some regions have coincident optical and X-ray emission, there are
other regions in the Cas A remnant that exhibit emission only in either optical
or X-ray bands. These regions do not seem to have obvious positional or 
temporal offsets
which might be attributed to ablation of dense ejecta knots. 

\begin{figure}
\includegraphics{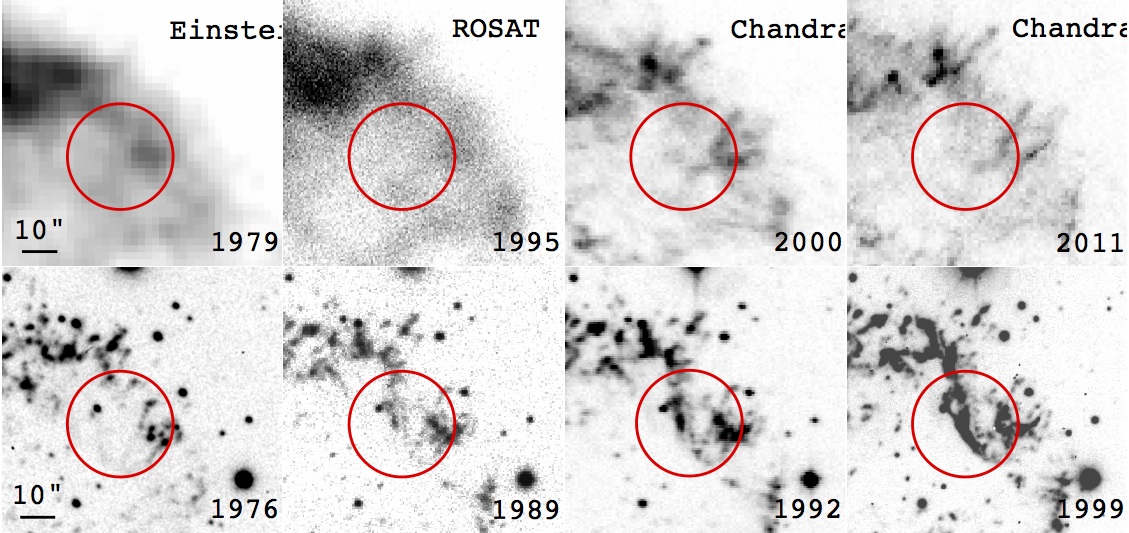}
\caption{Comparison of X-ray and optical emission for Cas A's northwestern rim.
The positions of the 30 arcsecond diameter red circles is the same on 
all X-ray and optical images.
{\it Upper panels}: X-ray images of a region marked by a red circle where a 
substantial increase
of optical filaments have appeared since the 1970s.
{\it Lower panels}: A series of red optical images shows the development of 
a bright band of curved filaments between 1976 and 1999.
The optical images shown are Palomar 5m images PH7252vB and PH8206vB,
an MDM 1.3m image taken using a broad [S II] interference filter,
and a MDM 2.4m image taken with an R filter.
Since 1999 and through 2011, this band of bright optical filaments 
has remained bright and has become
more extensive while in comparison only faint X-ray emission is seen in 
this region. }
\label{figure_NW_arc}
\end{figure}

\begin{figure}
\epsscale{0.75}
\includegraphics{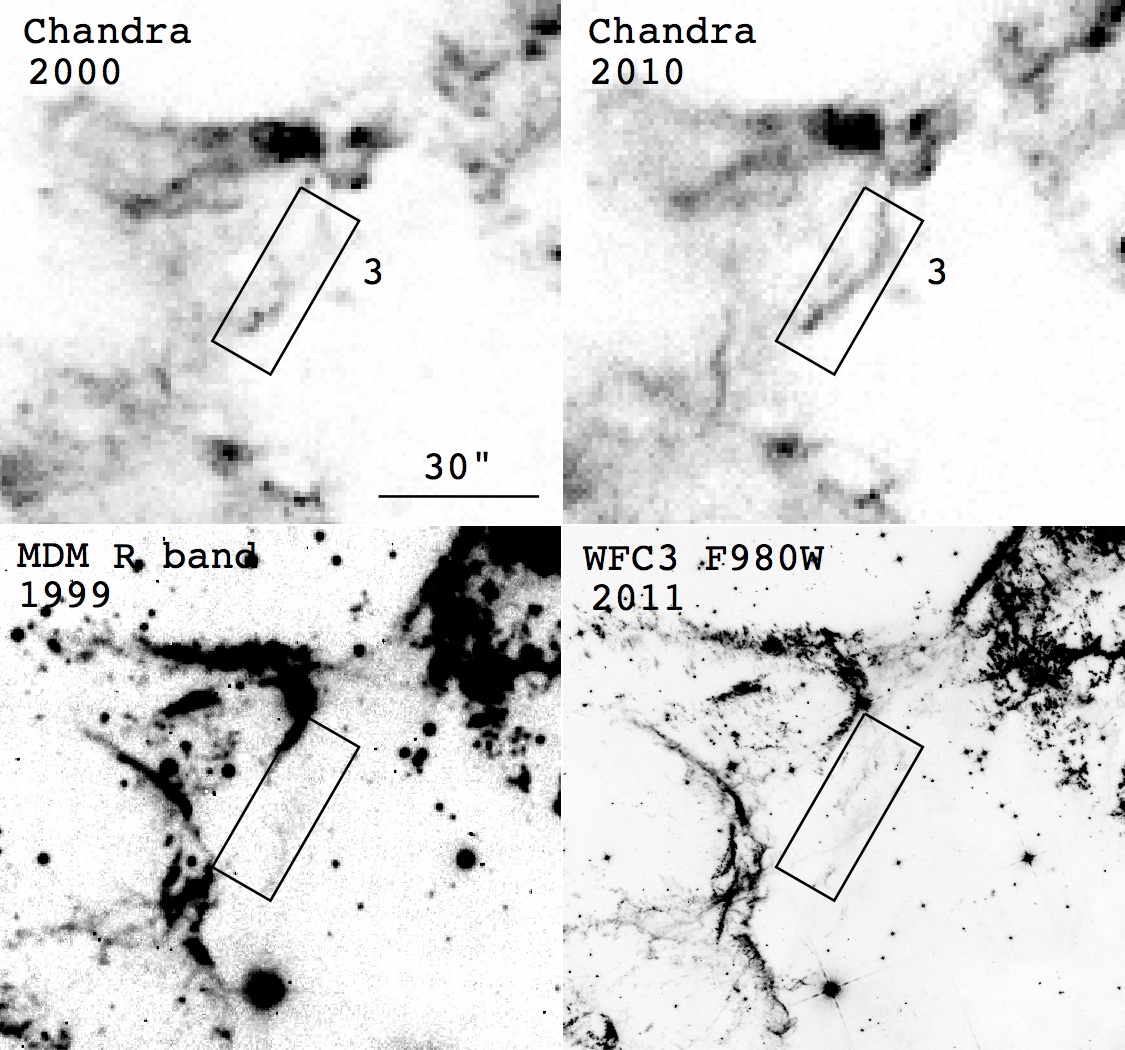}
\caption{X-ray and optical images of Cas A's eastern limb showing the
development of a bright X-ray filament with little associated optical 
emission.}
\label{fig:EastRegions}
\end{figure}

\begin{figure}
\epsscale{0.75}
\includegraphics{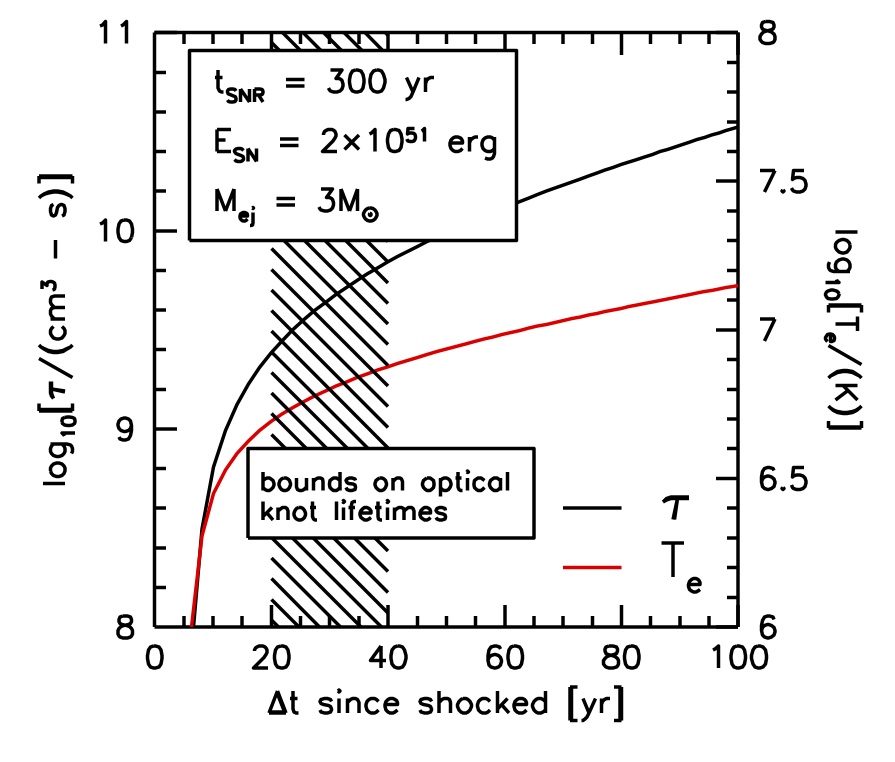}
\caption{Ionization parameter (black) and electron temperature (red) as a 
function of time since shocked. The model assumes an ejecta mass of 
3M$_{\sun}$,
explosion energy of 2$\times$10$^{51}$ erg, progenitor mass-loss rate of 
2$\times$10$^{-5}$ M$_{\sun}$ yr$^{-1}$ and a wind speed of 10 km s$^{-1}$.
Also shown as a hatched region are the bounds on optical knot fading, either
by radiative cooling or hydrodynamical destruction.}
\label{fig:model}
\end{figure}

\begin{figure}
\epsscale{0.75}
\includegraphics{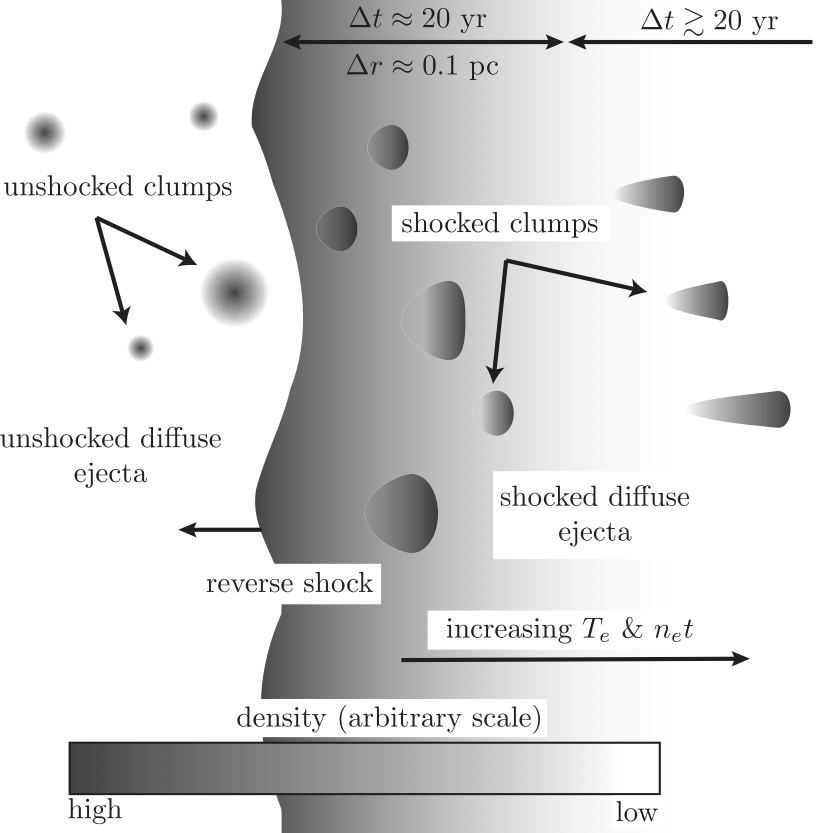}
\caption{Schematic representation of the structure of dense ejecta clumps. 
Dense clumps
with density $n_{\mathrm{c}}$ embedded in a more diffuse component with density
$n_{\mathrm{ej}}$ cross the reverse shock from left to right. The diffuse
ejecta is decelerated to three quarters the unshocked velocity, and a slow 
shock is driven into the clumps. While a shock is driven into the clumps, they
remain largely undecelerated by the reverse shock, due to their small cross 
section,
and behave in a ballistic fashion until they are destroyed by hydrodynamical
instabilities. At later times, the clumps have moved away from the
reverse shock, and while they begin to fade, the diffuse component becomes
bright in X-rays. These knots differ from those in Fig.~\ref{fig:opt_tails}
possibly due to the structure of the envelope which they are embedded in.}
\label{fig:ejecta}
\end{figure}

\begin{figure}
\epsscale{0.75}
\includegraphics{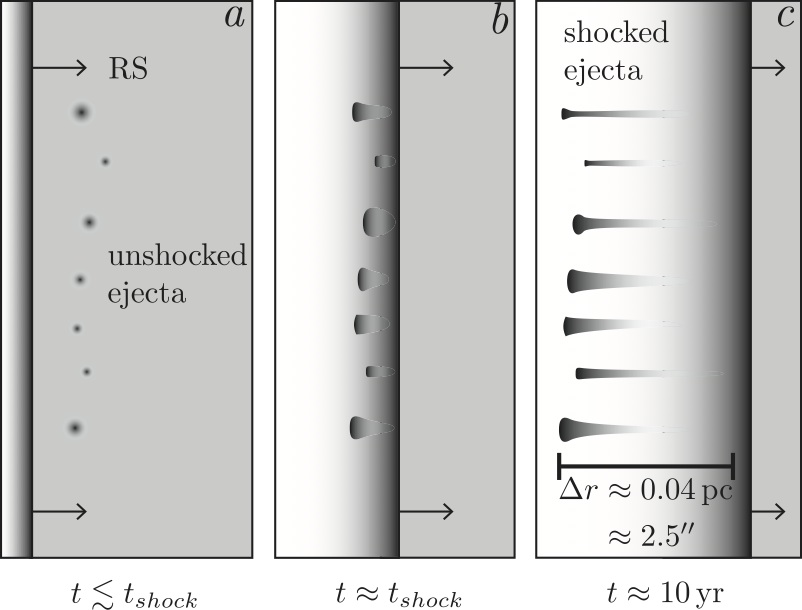}
\caption{Cartoon representation of dense knot ablation which leads to
dense optical knots positioned downstream from X-ray emission, which 
appears closer to the reverse shock. In panel {\it a}, the unshocked
knots are overtaken by the reverse shock. In panel {\it b}, a radiative
shock is driven into the knots, and material is ablated off the sides
of the knots, due to Kelvin--Helmholtz instabilities. In panel {\it c}, 
the ablated material appears as tails in optical emission, and as 
diffuse trailing emission in X-rays. Due to the lower velocity of the
ablated material, it remains closer to the reverse shock position.}
\label{fig:opt_tails}
\end{figure}

One such region lies in the remnant's northwest limb (Fig.~\ref{figure_NW_arc})
where the advance of Cas A's reverse shock front can be readily identified in
optical emission \citep{Morse2004}.  Between 1976 and 1992, the optical
emission from this region has brightened substantially. This is in contrast to
the X-ray emission (upper panels of Fig.~\ref{figure_NW_arc}), which show no
clearly discernible change between 1979 and 2011. \citet{Morse2004} estimated
preshock densities in this region of 25 -- 2500 cm$^{-3}$. Ram pressure
equilibrium thus requires that the velocity of this shocked high density
component to be $\approx$ 20--200 km s$^{-1}$. With these densities, it is not
surprising that there is no correlated X-ray/optical emission in this region
especially since it has evolved into a nearly continuous filament of optical
emission.

In contrast to the optically bright region in the northwest with no
corresponding X-ray emission, a completely opposite situation exists in the
northeast near the base of the remnant's NE jet. Shown in
Figure~\ref{fig:EastRegions} is a region of brightening X-ray emission with no
corresponding optical emission. Between 2000 and 2010, the filament has
brightened substantially, with little or no corresponding optical emission
seen. Using {\it Chandra} observations between 2000 and 2012, we modeled both
the Si-K 1.75--2.0 keV line flux and line centroid, and found that the flux has
increased by nearly a factor of 3 in 12 years, and the line centroid has
increased by 10 eV over the same period, indicating an ionizing plasma (see
Table~\ref{tab:xray_rs}).

While these two cases are quite different, the underlying
cause of the differences may again be attributed to ejecta density. In the case
of the X-ray emission in the northeast, lack of optical emission here could be
due to the ejecta density being too low ($\lesssim$ 1 cm$^{-3}$) resulting in a
high velocity of the shocked ejecta (probably several thousand kilometers per
second) thus making the conditions not conducive to the formation of optically
bright filaments. A similar situation may help explain the copious X-ray
emission in the east and southeast (Fig.~\ref{Southeast}) where one sees hardly
any coincident optical emission.  \citet{MF2013} noted the presence of plumes
and rings of ejecta in Cas A. In the east/southeast, the plume of ejecta here
might be of low density or at least any inhomogeneities might be smoothed out
by hydrodynamical effects and thus optical emission from dense knots is not
expected.

The northwest line of optical filaments shown in Figure~\ref{figure_NW_arc}
presents the exact opposite scenario as that seen in the southeast. In this
region, it is likely that the reverse shock is currently interacting with a
nearly continuous sheet of relatively high density material. The shocks driven
into this sheet are not strong enough to produce X-ray emission but does result
in a nearly continuous filament of optical emission. 

\subsection{A Clump + Envelope Model for Cas A's Ejecta}

A large fraction of Cas A's ejecta likely exists as a low density, diffuse
component with densities $\sim$ 1 cm$^{-3}$.  We suggest that embedded within
this low density, diffuse component are clumps of  varying densities, many with
surrounding lower density envelope-like structures.  Dense knots with
relatively thin envelopes (where the transition from the diffuse to clumped
ejecta is akin to a step function) give rise to a head--tail like structure
after reverse shock passage, resulting in the shocked optical knots appearing
downstream of associated X-ray emission which is located closer to the shock;
e.g., as in Region~1, shown in Figure~\ref{fig:Region1} or in the
``parentheses'' seen in Figure~\ref{fig:Parenthesis}, and idealized in the
schematic seen in Figure~\ref{fig:opt_tails}. 

There are also dense ejecta clumps that are surrounded by a much broader,
yet lower density, envelope (see Fig.~\ref{fig:region_a2e}).  In these cases,
the optical knots and X-ray emitting envelope remain co-spatial, though there
may be time delays as the optical emission brightens and fades, and then the
X-ray emission subsequently brightens. Finally, there are regions where there
is either only optical emission, and no X-ray emission, or visa versa (see
Figs.\ 16 - 18). Which type of emission one observes will mainly dependent upon
the region's density.

Taken together, these various emission morphologies point to the
ejecta being mostly diffuse in volume but containing small over-densities of
several orders of magnitude. The exact structure of the transition
region between the diffuse and clumped component is what determines
the time and spatial scales over which correlated emission is
observed. 

This ejecta structure model can also explain discrepancies in estimated
expansion ages for Cas ~A derived from various emission band observations. The
remnant's optical expansion rate is $\approx$ $0.3$\% yr$^{-1}$
\citep{Fesen2001,thor01}, while its X-ray expansion rate has been calculated to
be $0.2$\% yr$^{-1}$ \citep{koralesky98,vink98,delaney04}.  The diffuse X-ray
emission which gives rise to the lower expansion rate (and hence higher
expansion age) is more strongly decelerated by the reverse shock (5000 km
s$^{-1}$ undecelerated, to $\approx$ 3750 km s$^{-1}$ decelerated), while the
optical knots are hardly decelerated, 
remaining essentially ballistic. The X-ray
emitting material that is ablated off dense ejecta clumps (bullets) and 
exhibiting a
spatial offset from the optical emitting material is expected to
show a range of decelerations most similar to the diffuse component but not as
strongly decelerated leading to space motions not dissimilar from that of the
bullets.

\subsection{X-ray/Optical Correlations in Other Young SNRs}

Optical and X-ray emission correlations have been discussed in the context of
other young supernova remnants. Comparing [\ion{O}{3}] observations of SNR
G292.0+1.8 to an existing {\it Chandra} observation, \citet{ghavamian05} found
a good correlation between optical fast moving knots and more diffuse X-ray
emission.  Similar to our ejecta model, they postulated that the [\ion{O}{3}]
emission in G292.0+1.8 originated from radiative shocks driven into knots with
densities $\sim$ 10$^{3}$ cm$^{-3}$ while the X-ray emission arose from the
more diffuse interclump gas with densities of only a few.

Like G292.0+1.8, similar correspondences are observed in the Small Magellanic
Cloud SNR 1E 0102-7219.  \citet{gaetz00} noted that there were some
correlations between O-bright X-ray filaments and interior [\ion{O}{3}]
filaments observed with {\it HST} Planetary Camera F5202N, but noted that the
{\it HST} observations may have missed some high velocity [\ion{O}{3}] emission
that was Doppler-shifted out of the filter bandpass.  \citet{finkelstein06}
compared {\it HST} ACS observations of 1E 0102-7219 and found that the optical
emission and X-ray emission are not exactly coincident with each other. This is
consistent with the ejecta model presented here and by \citet{ghavamian05}
namely, optical emission is embedded or offset from X-ray emission due to
density effects and the finite time required to shock heat the ejecta to X-ray
emitting temperatures.

\section{Conclusions}

We have presented optical observations of Cassiopeia A dating back to 1951 and
up through 2011 and compare these data to X-ray observations taken
with {\it Einstein}, {\it ROSAT}, and {\it Chandra} in order to investigate
spatial correlations between X-ray and optical emissions. Due to the large
differences in postshock densities and temperatures, the prevailing view has
been that there is little correlation between ejecta that emits in
X-rays and that which emits optically. However, our study shows that this is
not always true. Taking into account the dynamical evolution of shocked ejecta
and the relevant X-ray and optical timescales involved in the radiative and
hydrodynamical evolution of shocked ejecta, we do find X-ray/optical
correlations in many regions of Cas A.

We have identified four cases of correlations and anti-correlations between
X-ray and optical emission in the shocked ejecta in Cas A. These are: 1) X-ray
and optical emission time delays of years or even several decades where the
optical emission for a region or feature shows up prior to its associated X-ray
emission, 2) spatial offsets typically of a few arcseconds ($\approx 10^{17}$)
cm) between a feature's optical and X-ray emission emission peaks, 3) regions
showing significant optical emission but with no corresponding X-ray emission,
and 4) strong X-ray emitting regions having little if any positional coincident
optical emission. 

To explain these correlations and anti-correlations, we propose a highly
inhomogeneous density model for Cas A's ejecta consisting of: 1) small dense
knots which rapidly form optical emission following reverse shock front passage
embedded in a more extended and more diffuse lower density component giving rise to
associated X-ray emission but sometimes showing a significant time delay
relative to the optical, 2) shock induced mass ablation off dense ejecta clumps
thereby generating trailing low density and X-ray emitted material and hence
positionally offset from the optical emitting knots, 3) smooth and relatively
continuous high density ejecta filaments or shell walls that never reach X-ray
emitting temperature, or conversely, large extended regions consisting mainly
of low density ejecta, both of which lead to X-ray/optical emission
anti-correlations. 

A highly inhomogeneous ejecta model as proposed here for Cas A may also help
explain some of the X-ray/optical emission features seen in other
young core collapse SN remnants. However it remains to be determined what
dynamical and radiative properties of the explosion mechanism sets the relative
volumes of a remnant's low and high density ejecta, how these components are
distributed and arranged in the expanding debris cloud, or what are the limits
to range of ejecta densities as a function of elemental abundances and
expansion velocity \citep[e.g.][]{kifonidis03}. 
Further studies into some of the properties of Cas~A's
interior unshocked ejecta like those reported by \citet{smith09},
\citet{isensee2010}, \citet{Grefen2014} and \citet{MF2014} may give us valuable
insights to these issues. 

\acknowledgements

We wish to thank Sidney van den Bergh for making available the extensive
Palomar plate collection of Cas A images dating back to 1951 and to Josh
Grindley and the DASCH plate scanning team at CfA/Harvard for their expert help
and assistance in scanning these priceless photographic plates.  We also thank
J.~Thorstensen for assistance with applying WCS coordinates to the scanned
Palomar plates. We thank Dan Milisavljevic, J.~Martin Laming, and Roger
Chevalier for useful comments on an early draft of this manuscript.  D.~J.~P.
acknowledges support from the {\it Chandra X-ray Center} through GO8-9065A and
from Space Telescope Science Institute through grant GO-11337.01-A. D.J.P.\
also acknowledges support through NASA contract NAS8-03060. R.A.F.\
acknowledges support from the National Science Foundation under grant
AST-0908237 and from NASA through grants GO-8281, 9238, 9890, 10286, 12300, and
12674 from the Space Telescope Science Institute, which is operated by the the
Association of Universities for Research in Astronomy, Inc.

\newpage

\begin{deluxetable}{lclcll}
\tablecolumns{6}
\tablewidth{0pc}
\tablecaption{Summary Log of Observations Used for this Analysis}
\tablehead{
\colhead{Instrument} & \colhead{Date} & 
\colhead{Obs. ID} & \colhead{Exposure Time}   & \colhead{Energy/Wavelength\tablenotemark{a}} & Observer/PI   }
\startdata
\underline{\bf{X-Ray} ~~~~~~~~~~}       &            &          &          & \nodata  &       \\
Einstein HRI                & 1979-02-08 & 713      & 42.5 ks    & 0.1--4.5 keV         & Seward          \\
ROSAT HRI                         & 1995-12-23 &US500444H & 175 ks   & 0.2--2.0 keV        & Keohane  \\
{\it CXO} ACIS-S\tablenotemark{b} & 2000-01-30 & 00114 & 50.0 ks   & 0.3--10 keV      & Holt         \\
{\it CXO} ACIS-S\tablenotemark{b} & 2002-02-06 & 01952 & 50.0 ks   & 0.3--10 keV      & Rudnick         \\
{\it CXO} ACIS-S\tablenotemark{b} & 2004-02-08 & 05196 & 50.0 ks   & 0.3--10 keV      & Hwang          \\
{\it CXO} ACIS-S\tablenotemark{b} & 2007-12-05 & 09117 & 25.0 ks   & 0.3--10 keV       & Patnaude         \\
{\it CXO} ACIS-S\tablenotemark{b} & 2007-12-08 & 09773 & 25.0 ks   & 0.3--10 keV       & Patnaude  \\
{\it CXO} ACIS-S\tablenotemark{b} & 2009-11-02 & 10935 & 25.0 ks   & 0.3--10 keV       & Patnaude  \\
{\it CXO} ACIS-S\tablenotemark{b} & 2009-11-03 & 12020 & 25.0 ks   & 0.3--10 keV       & Patnaude  \\
{\it CXO} ACIS-S\tablenotemark{b} & 2010-10-31 & 10936 & 31.0 ks   & 0.3--10 keV       & Patnaude  \\
{\it CXO} ACIS-S\tablenotemark{b} & 2010-11-02 & 13177 & 19.0 ks   & 0.3--10 keV       & Patnaude  \\
{\it CXO} ACIS-S\tablenotemark{b} & 2012-05-15 & 14229 & 49.0 ks   & 0.3--10 keV       & Patnaude  \\
{\it CXO} ACIS-S\tablenotemark{b} & 2013-05-20 & 14480 & 49.0 ks   & 0.3--10 keV       & Patnaude  \\
\underline{\bf{Optical} ~~~~~~~~~}         &                &              &    &                \\
Palomar Hale 5m & 1951-09-09 & PH520B        & 1800 s       & blue: 103aO+GG1         & Baade      \\
Palomar Hale 5m & 1951-11-01 & PH553B        & 7200 s       & ~red: 103aE+RG2         & Baade      \\
Palomar Hale 5m & 1951-11-03 & PH563B        & 7200 s       & ~red: 103aE+RG2         & Baade      \\
Palomar Hale 5m & 1953-08-11 & PH778B        & 1500 s       & blue: 103aO+GG1         & Baade      \\
Palomar Hale 5m & 1953-08-13 & PH793B        & 6120 s       & ~red: 103aE+RG2         & Baade      \\
Palomar Hale 5m & 1954-11-25 & PH232M        & 5400 s       & blue: 103aJ+GG11        & Minkowski      \\
Palomar Hale 5m & 1954-11-26 & PH236M        & 7200 s       & ~red: 103aE+OR1         & Minkowski  \\
Palomar Hale 5m & 1954-12-23 & PH1159B       & 1800 s       & blue: 103aO+GG1         & Baade \\
Palomar Hale 5m & 1957-09-21 & PH1732B       & 5400 s       & ~red: 103aF+RG2         & Baade    \\
Palomar Hale 5m & 1957-09-22 & PH1736B       & 2100 s       & blue: 103aO+GG1         & Baade   \\
Palomar Hale 5m & 1958-08-11 & PH3033S       & 5400 s       & ~red: 103aF+RG2         & Shane     \\
Palomar Hale 5m & 1965-08-30 & PH4815        & 5400 s       & ~red: 103aE+0R1         & Miller  \\
Palomar Hale 5m & 1967-10-04 & PH5107A       & 4200 s       & ~red: 103aE+RG2         & Arp    \\
Palomar Hale 5m & 1968-09-26 & PH5254vB      & 5400 s       & ~red: 103aF+RG2         & van den Bergh \\
Palomar Hale 5m & 1968-09-26 & PH5255vB      & 1500 s       & blue: 103aD+GG11        & van den Bergh \\
Palomar Hale 5m & 1970-09-01 & PH5643vB      & 7200 s       & blue: 103aJ+GG475       & van den Bergh  \\
Palomar Hale 5m & 1970-09-03 & PH5648vB      & 7200 s       & blue: 103aJ+GG475       & van den Bergh \\
Palomar Hale 5m & 1970-09-04 & PH5659vB      & 7200 s       & ~red: 103aE+RG630       & van den Bergh  \\  
Palomar Hale 5m & 1971-08-29 & PH5946        & 5400 s       & ~red: 103aF+RG630       & Arp            \\
Palomar Hale 5m & 1972-09-09 & PH6236vB      & 7200 s       & blue: IIIaJ+GG7         & van den Bergh \\
Palomar Hale 5m & 1972-09-10 & PH6249vB      & 7200 s       & ~red: 103aF+RG2         & van den Bergh \\
Palomar Hale 5m & 1973-07-31 & PH6555vB      & 3600 s       & ~red: 098-04+RG2        & van den Bergh \\
Palomar Hale 5m & 1973-08-01 & PH6562vB      & 7200 s       & blue: 103aJ+GG7         & van den Bergh \\
Palomar Hale 5m & 1973-08-04 & PH6573vB      & 12000 s      & 098+[S II](FWHM=166 A)  & van den Bergh \\
Palomar Hale 5m & 1973-08-13 & PH6891vB      & 4200 s       & ~red: 098-04+RG2        & van den Bergh \\
Palomar Hale 5m & 1974-08-14 & PH6902vB      & 7200 s       & blue: IIIaJ+GG7         & van den Bergh \\
Palomar Hale 5m & 1974-08-15 & PH6914vB      & 7200 s       & blue: 127-02+GG7        & van den Bergh \\
Palomar Hale 5m & 1975-07-18 & PH7077vB      & 6000 s       & ~red: 098-02+RG645      & van den Bergh  \\
Palomar Hale 5m & 1976-06-30 & PH7231vB      & 7200 s       & blue: 127-04+GG7        & van den Bergh \\
Palomar Hale 5m & 1976-07-02 & PH7252vB      & 7200 s       & ~red: 098-04+RG645      & van den Bergh \\
Palomar Hale 5m & 1977-10-07 & PH7425vB      & 5400 s       & blue: 124-01+GG7        & van den Bergh \\
Palomar Hale 5m & 1977-10-08 & PH7433vB      & 4500 s       & ~red: 098-04+RG2        & van den Bergh \\
Palomar Hale 5m & 1977-10-10 & PH7445vB      & 7200 s       & blue: 103aJ+GG7         & van den Bergh \\
Palomar Hale 5m & 1980-07-13 & PH7766vB      & 6000 s       & ~red: 098-04+RG645      & van den Bergh \\
Palomar Hale 5m & 1980-07-14 & PH7771vB      & 7200 s       & blue: IIIaJ+GG7         & van den Bergh \\
Palomar Hale 5m & 1983-07-12 & PH8192vB      & 6000 s       & ~red: 098+RG645         & van den Bergh \\
Palomar Hale 5m & 1983-07-13 & PH8199vB      & 8700 s       & blue: 124-01:GG7        & van den Bergh \\
Palomar Hale 5m & 1983-07-14 & PH8201vB      & 10800 s      & 098+[S II](FWHM=166 A)  & van den Bergh \\
Palomar Hale 5m & 1989-09-28 & PH8202vB      & 7200 s       & ~red: 098-04+RG645      & van den Bergh \\
Palomar Hale 5m & 1989-09-28 & PH8204vB      & 7200 s       & 098+[S II](FWHM=166 A)  & van den Bergh \\
Palomar Hale 5m & 1989-09-29 & PH8206vB      & 7200 s       & ~red: 098-04+RG645      & van den Bergh \\
Palomar Hale 5m & 1989-09-29 & PH8207vB      & 7200 s       & blue: 103aJ+GG7         & van den Bergh \\
MDM 1.3m        & 1992-07-04 & ~~~\nodata    & 1000 s       & 6600--6800 \AA: broad [\ion{S}{2}]  & Fesen \\
MDM 2.4m        & 1998-09-18 & ~~~\nodata    & 2400 s       & 5600--8800 \AA: R band  & Fesen  \\
MDM 2.4m        & 1999-10-15 & ~~~\nodata    & 2400 s       & 5600--8800 \AA: R band  & Thorstensen  \\
{\it HST} WFPC2 & 2000-01-23 & U59T0*     & 1000 s       & 6000--7500 \AA: F675W   & Fesen  \\
{\it HST} WFPC2 & 2002-01-09 & U6D10*     & 1000 s       & 6000--7500 \AA: F675W   & Fesen  \\
{\it HST} ACS/WFC & 2004-12-05 & J8ZM0*   & 2400 s       & 5500--7100 \AA: F625W   & Fesen  \\
{\it HST} ACS/WFC & 2004-12-05 & J8ZM0*   & 2000 s       & 6900--8600 \AA: F775W   & Fesen  \\
{\it HST} WFPC2   & 2008-02-05 & UA5K0*   & 2400 s       & 6000--7500 \AA: F675W   & Patnaude  \\
{\it HST} WFPC2   & 2008-06-02 & UA5K0*   & 2400 s       & 6000--7500 \AA: F675W   & Patnaude  \\
{\it HST} WFC3/IR & 2010-10-28 & IBID*   & 22.1 ks      & 9000--10700 \AA: F098M  & Fesen  \\
{\it HST} WFC3/IR & 2011-11-18 & IBQH*   & 22.1 ks      & 9000--10700 \AA: F098M  & Fesen 
\enddata
\tablenotetext{a}{For the Palomar observations, we include the
photographic emulsion and filter used.}
\tablenotetext{b}{{\it Chandra} ACIS images are broadband images, but as
noted in the text, they have been energy filtered, fluxed, and continuum
subtracted for the purposes of comparison with the {\sl Einstein} and ROSAT HRI images.}
\label{tab:obs}
\end{deluxetable}

\begin{deluxetable}{lccccc}
\tablecolumns{6}
\tablewidth{0pc}
\tablecaption{Estimated 0.1-4.5 keV Absorbed X-ray Fluxes: 1979--2013}
\tablehead{
\colhead{Observatory} & \colhead{Feature A\tablenotemark{a}} & 
\colhead{Feature B} & \colhead{Feature C} & \colhead{Feature D} & 
\colhead{Feature E} }
\startdata
{\it Einstein} HRI (1979)   & 7.3$\pm$0.8  & 0.0$\pm$0.3  & 11.0$\pm$1.2 & 5.6$\pm$0.6  & 14.0$\pm$1.7 \\
{\it ROSAT} HRI (1995)      & 9.4$\pm$1.1  & 5.5$\pm$0.5  & 11.0$\pm$0.9 & 12.0$\pm$1.3 & 14.0$\pm$1.3 \\
{\it Chandra} ACIS-S (2000) & 23.6$\pm$3.1 & 11.0$\pm$1.0 & 31.7$\pm$3.8 & 19.0$\pm$1.8 & 32.2$\pm$3.6 \\
{\it Chandra} ACIS-S (2004) & 26.4$\pm$2.1 & 14.1$\pm$1.3 & 32.7$\pm$3.3 & 23.9$\pm$2.2 & 39.9$\pm$4.0 \\
{\it Chandra} ACIS-S (2007) & 28.9$\pm$3.1 & 16.8$\pm$1.5 & 43.9$\pm$4.5 & 28.9$\pm$2.7 & 42.1$\pm$3.8 \\
{\it Chandra} ACIS-S (2013) & 40.0$\pm$3.8 & 23.8$\pm$2.5 & 57.4$\pm$5.5 & 32.9$\pm$3.6 & 49.1$\pm$5.1 
\enddata
\tablenotetext{a}{X-ray fluxes are given in units of 10$^{-13}$ erg cm$^{-2}$
s$^{-1}$}
\label{tab:xrayflux}
\end{deluxetable}

\begin{deluxetable}{lcc}
\tablecolumns{6}
\tablewidth{0pc}
\tablecaption{Si-K line centroid and 1.75 -- 2.0 keV modeled flux
for northeast filament at the base of the ``jet'': 2000--2012}
\tablehead{
\colhead{Epoch} & \colhead{Line Centroid} & \colhead{Modeled Flux} \\
\colhead{} & \colhead{keV} & \colhead{10$^{-13}$ erg cm$^{-2}$ s$^{-1}$}}
\startdata
2000 & 1.852$\pm$0.003 & 2.9 \\
2004 & 1.858$\pm$0.003 & 4.0 \\
2008 & 1.862$\pm$0.002 & 6.3 \\
2012 & 1.863$\pm$0.002 & 8.7 
\enddata
\label{tab:xray_rs}
\end{deluxetable}

\clearpage


\clearpage


\clearpage


\begin{thebibliography}

\bibitem[Araya et al.(2010)]{araya10} Araya, M., Lomiashvili, 
         D., Chang, C., Lyutikov, M., \& Cui, W.\ 2010, \apj, 714, 396 

\bibitem[Baade \& Minkowski(1954)]{BM54} Baade, W., \& Minkowski, R.\ 1954, \apj, 119, 206 

\bibitem[Besel \& Krause(2012)]{Besel12} Besel, M.-A., \& Krause, O.\ 2012, \aap, 541, L3 

\bibitem[Blondin \& Lufkin(1993)]{blondin93} Blondin, J.~M., \& Lufkin, E.~A.\ 
         1993, \apjs, 88, 589 

\bibitem[Borkowski et al.(1989)]{borkowski89} Borkowski, K.~J., 
         Shull, J.~M., \& McKee, C.~F.\ 1989, \apj, 336, 979 

\bibitem[Chevalier \& Kirshner(1978)]{ck78} Chevalier, R. A., \& Kirshner,
          R. P. 1978, \apj, 219, 931

\bibitem[Chevalier \& Kirshner(1979)]{ck79} Chevalier, R. A., \& Kirshner,
         R. P. 1979, \apj, 233, 154

\bibitem[Chevalier \& Oishi(2003)]{chevalier03} Chevalier, R.~A., \& Oishi, J.\ 
         2003, \apjl, 593, L23 

\bibitem[Claeys et al.(2011)]{Claeys11} Claeys, J.~S.~W.,
         de Mink, S.~E., Pols, O.~R.,
          Eldridge, J.~J., \& Baes, M.\ 2011, \aap, 528, A131

\bibitem[DeLaney \& Rudnick(2003)]{delaney03} DeLaney, T., \& Rudnick, L.\ 
         2003, \apj, 589, 818 

\bibitem[DeLaney et al.(2004)]{delaney04} DeLaney, T., Rudnick, 
         L., Fesen, R.~A., et al.\ 2004, \apj, 613, 343 

\bibitem[DeLaney et al.(2010)]{delaney10} DeLaney, T., et al.\ 
         2010, \apj, 725, 2038 

\bibitem[DeLaney et al.(2014)]{delaney14} DeLaney, T., Kassim, 
N.~E., Rudnick, L., \& Perley, R.~A.\ 2014, arXiv:1403.0032 

\bibitem[Eriksen et al.(2009)]{eriksen09} Eriksen, K.~A., Arnett, 
D., McCarthy, D.~W., \& Young, P.\ 2009, \apj, 697, 29 

\bibitem[Fabian et al.(1980)]{fabian80} Fabian, A.~C., 
         Willingale, R., Pye, J.~P., Murray, S.~S., 
         \& Fabbiano, G.\ 1980, \mnras, 193, 175 

\bibitem[Fesen(2001)]{Fesen2001}Fesen, R. A. 2001, \apjs, 133, 161

\bibitem[Fesen et al.(2001)]{Fesenetal2001} Fesen, R.~A., Morse, 
         J.~A., Chevalier, R.~A., et al.\ 2001, \aj, 122, 2644 

\bibitem[Fesen et al.(2006)]{fesen06} Fesen, R.~A., et al.\ 
         2006, \apj, 645, 283 

\bibitem[Fesen et al.(2011)]{Fesen11} Fesen, R.~A., Zastrow, 
         J.~A., Hammell, M.~C., Shull, J.~M., \& Silvia, D.~W.\ 2011, \apj, 736, 109 

\bibitem[Finkelstein et al.(2006)]{finkelstein06} Finkelstein, S.~L., 
Morse, J.~A., Green, J.~C., et al.\ 2006, \apj, 641, 919 

\bibitem[Gaetz et al.(2000)]{gaetz00} Gaetz, T.~J., Butt, 
Y.~M., Edgar, R.~J., et al.\ 2000, \apjl, 534, L47 

\bibitem[Ghavamian et al.(2005)]{ghavamian05} Ghavamian, P., 
Hughes, J.~P., \& Williams, T.~B.\ 2005, \apj, 635, 365 

\bibitem[Giacconi et al.(1979)]{giacconi79} Giacconi, R., 
         Branduardi, G., Briel, U., et al.\ 1979, \apj, 230, 540 

\bibitem[Grefenstette et al.(2014)]{Grefen2014} Grefenstette, 
         B.~W., Harrison, F.~A., Boggs, S.~E., et al.\ 2014, \nat, 506, 339

\bibitem[Hamilton \& Sarazin(1984)]{hamilton84} Hamilton, A.~J.~S., \& 
         Sarazin, C.~L.\ 1984, \apj, 287, 282 

\bibitem[Hughes et al.(2000)]{hughes00} Hughes, J.~P., Rakowski, 
         C.~E., Burrows, D.~N., \& Slane, P.~O.\ 2000, \apjl, 528, L109 

\bibitem[Hurford \& Fesen(1996)]{Hurford96} Hurford, A.~P., \& Fesen, R.~A.\ 1996, \apj, 469, 246 

\bibitem[Hwang et al.(2000)]{Hwang00} Hwang, U., Holt, S.~S., \& Petre, R.\ 2000, \apjl, 537, L119 

\bibitem[Hwang \& Laming(2003)]{hwang03} Hwang, U., \& Laming, J.~M.\ 2003, \apj, 597, 362 

\bibitem[Hwang et al.(2004)]{hwang04} Hwang, U., Laming, J.~M., 
Badenes, C., et al.\ 2004, \apjl, 615, L117 

\bibitem[Hwang \& Laming(2009)]{hwang09} Hwang, U., \& Laming, J.~M.\ 2009, \apj, 703, 883 

\bibitem[Hwang \& Laming(2012)]{hwang12} Hwang, U., \& Laming, J.~M.\ 2012, \apj, 746, 130 

\bibitem[Isensee et al.(2010)]{isensee2010} Isensee, K., Rudnick, 
         L., DeLaney, T., et al.\ 2010, \apj, 725, 2059 

\bibitem[Kifonidis et al.(2003)]{kifonidis03} Kifonidis, K., Plewa, T., 
Janka, H.-T., \& M\"{u}ller, E.\ 2003, \aap, 408, 621 

\bibitem[Klein et al.(1994)]{klein94} Klein, R.~I., McKee, 
         C.~F., \& Colella, P.\ 1994, \apj, 420, 213 

\bibitem[Koralesky et al.(1998)]{koralesky98} Koralesky, B., 
        Rudnick, L., Gotthelf, E.~V., \& Keohane, J.~W.\ 1998, \apjl, 505, L27 

\bibitem[Krause et al.(2008)]{krause08} Krause, O., Birkmann, 
         S.~M., Usuda, T., Hattori, T., Goto, M., Rieke, G.~H., 
         \& Misselt, K.~A.\ 2008, Science, 320, 1195 

\bibitem[Laming \& Hwang(2003)]{laming03} Laming, J.~M., \& Hwang, U.\ 2003, \apj, 597, 347 

\bibitem[Laming et al.(2006)]{laming06} Laming, J.~M., Hwang, 
         U., Radics, B., Lekli, G., \& Tak{\'a}cs, E.\ 2006, \apj, 644, 260 

\bibitem[Lawrence et al.(1995)]{law95} Lawrence, S.~S., 
         MacAlpine, G.~M., Uomoto, A., et al.\ 1995, \aj, 109, 2635 

\bibitem[Laycock et al.(2010)]{Laylock2010} Laycock, S., Tang, S., 
         Grindlay, J., et al.\ 2010, \aj, 140, 1062 

\bibitem[Markert et al.(1983)]{markert83} Markert, T.~H., Clark, 
         G.~W., Winkler, P.~F., \& Canizares, C.~R.\ 1983, \apj, 268, 134 

\bibitem[Masai(1994)]{masai94} Masai, K.\ 1994, \apj, 437, 770 

\bibitem[Milisavljevic \& Fesen(2013)]{MF2013} Milisavljevic, D., \& Fesen, R.~A.\ 2013, \apj, 772, 134 

\bibitem[Milisavljevic \& Fesen(2014)]{MF2014} Milisavljevic, D., \& Fesen, R.~A., 2013,
         in preparation

\bibitem[Minkowski(1959)]{minkowski59} Minkowski, R.\ 1959, URSI 
         Symp.~1: Paris Symposium on Radio Astronomy, 9, 315 

\bibitem[Morse et al.(2004)]{Morse2004} Morse, J.~A., Fesen, 
         R.~A., Chevalier, R.~A., et al.\ 2004, \apj, 614, 727 

\bibitem[Orlando et al.(2005)]{orlando05} Orlando, S., Peres, G., 
Reale, F., et al.\ 2005, \aap, 444, 505 

\bibitem[Patnaude \& Fesen(2007)]{patnaude07} Patnaude, D.~J., \& Fesen, R.~A.\ 2007, \aj, 133, 147 

\bibitem[Patnaude \& Fesen(2009)]{patnaude09a} Patnaude, D.~J., \& Fesen, R.~A.\ 2009, \apj, 697, 535 

\bibitem[Patnaude et al.(2009)]{patnaude09b} Patnaude, D.~J., 
         Ellison, D.~C., \& Slane, P.\ 2009, \apj, 696, 1956 

\bibitem[Patnaude et al.(2011)]{patnaude11} Patnaude, D.~J., Vink, 
         J., Laming, J.~M., \& Fesen, R.~A.\ 2011, \apjl, 729, L28 

\bibitem[Reed et al.(1995)]{Reed1995} Reed, J.~E., Hester, 
        J.~J., Fabian, A.~C., \& Winkler, P.~F.\ 1995, \apj, 440, 706 

\bibitem[Rest et al.(2008)]{rest08} Rest, A., et al.\ 2008, \apjl, 681, L81 

\bibitem[Rest et al.(2011)]{rest11} Rest, A., et al.\ 2011, \apj, 732, 3 

\bibitem[Ryle \& Smith(1948)]{Ryle1948} Ryle, M., \& Smith, F.~G.\ 1948, \nat, 162, 462 

\bibitem[Schure et al.(2008)]{schure08} Schure, K.~M., Vink, J., 
        Garc{\'{\i}}a-Segura, G., \& Achterberg, A.\ 2008, \apj, 686, 399 

\bibitem[Simcoe et al.(2006)]{Simcoe2006} Simcoe, R.~J., Grindlay, 
         J.~E., Los, E.~J., et al.\ 2006, \procspie, 6312,  

\bibitem[Smith et al.(2009)]{smith09} Smith, J.~D.~T., Rudnick, 
         L., Delaney, T., Rho, J., Gomez, H., Kozasa, T., Reach, W., 
         \& Isensee, K.\ 2009, \apj, 693, 713 

\bibitem[Sutherland 
\& Dopita(1995)]{sutherland95} Sutherland, R.~S., \& Dopita, M.~A.\ 1995, 
\apj, 439, 381 

\bibitem[Thorstensen et al.(2001)]{thor01} Thorstensen, J.~R., 
Fesen, R.~A., \& van den Bergh, S.\ 2001, \aj, 122, 297 

\bibitem[Truelove 
\& McKee(1999)]{truelove99} Truelove, J.~K., \& McKee, C.~F.\ 1999, \apjs, 120, 299 

\bibitem[van den Bergh(1971)]{vdb71a} van den Bergh, S.\ 1971, \apj, 165, 259 

\bibitem[van den Bergh(1971)]{vdb71b} van den Bergh, S. 1971, \apj, 165, 457

\bibitem[van den Bergh \& Dodd(1970)]{vdbD70} van den Bergh, S., \& Dodd, W. W. 1970,
          \apj, 162, 485

\bibitem[van den Bergh 
\& Kamper(1985)]{vandenbergh85} van den Bergh, S., \& Kamper, K.\ 1985, \apj, 293, 537 

\bibitem[Vink et 
al.(1996)]{vink96} Vink, J., Kaastra, J.~S., \& Bleeker, J.~A.~M.\ 1996, \aap, 307, L41 

\bibitem[Vink et 
al.(1998)]{vink98} Vink, J., Bloemen, H., Kaastra, J.~S., \& Bleeker, J.~A.~M.\ 1998, \aap, 339, 201 

\bibitem[Willingale et 
al.(2002)]{willingale02} Willingale, R., Bleeker, J.~A.~M., van der Heyden, 
K.~J., Kaastra, J.~S., \& Vink, J.\ 2002, \aap, 381, 1039 

\bibitem[Willingale et 
al.(2003)]{willingale03} Willingale, R., Bleeker, J.~A.~M., van der Heyden, 
K.~J., \& Kaastra, J.~S.\ 2003, \aap, 398, 1021 

\bibitem[Wongwathanarat et 
al.(2013)]{wong13} Wongwathanarat, A., Janka, H.-T., M\"{u}ller, 
E.\ 2013, \aap, 552, A126 

\bibitem[Young et al.(2006)]{young06} Young, P.~A., et al.\ 
2006, \apj, 640, 891 

\bibitem[Zacharias et al.(2010)]{Zach2010} Zacharias, N., Finch, 
         C., Girard, T., et al.\ 2010, \aj, 139, 2184 

\end{thebibliography}
\end{document}